\documentclass[apj]{emulateapj}

\shorttitle{Radio imaging observations of PSR J1023+0038}

\newcommand{\Msun}{\ensuremath{M_{\odot}}}

\begin{document}

\title{Radio imaging observations of PSR J1023+0038 in an LMXB state}

\author{A. T. Deller\altaffilmark{1}, J. Moldon\altaffilmark{1}, J. C. A. Miller-Jones\altaffilmark{2}, A. Patruno\altaffilmark{3,1}, J. W. T. Hessels\altaffilmark{1,4}, 
A. M. Archibald\altaffilmark{1}, Z. Paragi\altaffilmark{5}, G. Heald\altaffilmark{1,6}, N. Vilchez\altaffilmark{1}}
\altaffiltext{1}{ASTRON, the Netherlands Institute for Radio Astronomy, Postbus 2, 7990 AA, Dwingeloo, The Netherlands}
\altaffiltext{2}{International Centre for Radio Astronomy Research, Curtin University, GPO Box U1987, Perth, WA 6845, Australia}
\altaffiltext{3}{Leiden Observatory, Leiden University, P.O. Box 9513, 2300RA Leiden, The Netherlands}
\altaffiltext{4}{Anton Pannekoek Institute for Astronomy, University of Amsterdam, Science Park 904, 1098 XH Amsterdam, The Netherlands}
\altaffiltext{5}{Joint Institute for VLBI ERIC, Postbus 2, 7990AA Dwingeloo, The Netherlands}
\altaffiltext{6}{Kapteyn Astronomical Institute, University of Groningen, PO Box 800, 9700 AV, Groningen, The Netherlands}

\begin{abstract}
The transitional millisecond pulsar binary system PSR J1023+0038 re-entered an accreting state in 2013 June, in which it bears many similarities to low-mass X-ray binaries (LMXBs) in quiescence or near-quiescence.  At a distance of just 1.37~kpc, PSR J1023+0038 offers an unsurpassed ability to study low-level accretion onto a highly magnetized compact object.  We have monitored PSR J1023+0038 intensively using radio imaging with the Karl G. Jansky Very Large Array, the European VLBI Network and LOFAR, seeing rapidly variable, flat spectrum emission that persists over a period of six months.  The flat spectrum and variability are indicative of synchrotron emission originating in an outflow from the system, most likely in the form of a compact, partially self-absorbed jet, as is seen in LMXBs at higher accretion rates.  The radio brightness, however, greatly exceeds extrapolations made from observations of more vigorously accreting neutron star LMXB systems.  We postulate that PSR J1023+0038 is undergoing radiatively inefficient `propeller-mode' accretion, with the jet carrying away a dominant fraction of the liberated accretion luminosity. We confirm that the enhanced $\gamma$-ray emission seen in PSR J1023+0038 since it re-entered an accreting state has been maintained; the increased $\gamma$-ray emission in this state can also potentially be associated with propeller-mode accretion. Similar accretion modes can be invoked to explain the radio and X-ray properties of the other two known transitional millisecond pulsar systems XSS J12270-4859 and PSR J1824-2452I (M28I), suggesting that radiatively inefficient accretion may be an ubiquitous phenomenon amongst (at least one class of) neutron star binaries at low accretion rates.
\end{abstract}


\section{Introduction}

Pulsars are rotating neutron stars, spinning down as they lose their rotational energy in the form of a pair-plasma wind, $\gamma$- and X-rays, and radio waves. Millisecond pulsars (MSPs), with periods $\lesssim 20$ ms and fields $\sim 10^8$ G, have long been thought to be old pulsars spun up by a ``recycling'' process \citep{smarr76a,alpar82a,radhakrishnan82a} in low-mass X-ray binaries (LMXBs). During this process, material from a lighter companion is transferred to the pulsar via an accretion disk, spinning it up. This general model is well-supported by observations of neutron star LMXBs, which in a few cases actually show millisecond X-ray pulsations as the accreting material forms hotspots at the magnetic poles of the neutron star \citep[the so-called accreting millisecond X-ray pulsars or AMXPs;][and references therein]{wijnands98a,patruno12a}. That said, the end of this recycling process, when accretion stops and the radio pulsar becomes active, remains mysterious: why and when does accretion stop? How does the episodic nature of LMXB accretion affect the spin frequency of the neutron star? What happens when the accretion rates are too low to overcome the ``centrifugal barrier'' imposed by the pulsar's magnetic field and spin? What happens to the accretion stream when the pulsar becomes active, producing a powerful pair-plasma wind? What is the ultimate fate of the companion in such a system?

A range of ideas have been put forward to answer these questions, but they have been very difficult to test against observations due to the lack of systems known to be in such transitional states. Recently, however, three systems have been observed to transition between an (eclipsing) radio pulsar state and a LMXB state where an accretion-disk is present: PSR~J1023+0038 \citep[hereafter J1023;][]{archibald09a,patruno14a,stappers14a}, PSR J1824-2452I \citep[M28I;][]{papitto13a,linares14a}, and XSS~J12270$-$4859 \citep[hereafter XSS~J12270;][]{bassa14a,roy15a}.  A fourth system, 1RXS~J154439.4$-$112820, is also suspected to be a member of this same class of transitional systems, but has not yet been seen as a radio pulsar \citep{bogdanov15a}.

It is already clear that the end of accretion is an episodic process.  During the episodes where an accretion disk is present, several possible inner-disk behaviours are suggested by X-ray observations.  M28I exhibited millisecond X-ray pulsations and thermonuclear bursts at relatively high X-ray luminosities ($L_X \sim 10^{36}$ erg s$^{-1}$), attributes consistent with AMXPs during outburst.  J1023 and XSS~J12270, however, have never been observed by X-ray all-sky monitors to exceed $L_X \sim 10^{34.5}$ erg s$^{-1}$, and the nature of these relatively faint X-ray states is unclear. One possibility would be so-called ``propeller-mode accretion'', in which material from the companion forms an accretion disk extending down into the pulsar's light cylinder and shorting out the radio pulsar activity \citep{illarionov75a,ustyugova06a}. In propeller-mode accretion, the pressure of infalling material is balanced by the magnetic field of the neutron star outside the corotation radius; as a result, the neutron star's rotation accelerates the inner edge of the disk and material cannot fall further inward (the ``centrifugal barrier''), instead being ejected in polar outflows.  If the gas cannot be expelled from the system \citep[for example, because the centrifugal barrier is insufficient to bring the gas to the escape velocity;][]{spruit93a} then the inner accretion disk regions can become a ``dead disk" \citep{syunyaev77a} and remain trapped near co-rotation \citep{dangelo10a,dangelo12a}.  In this case episodic accretion onto the neutron-star is still possible.  \citet{eksi05a} suggest that there may even be a narrow window in which a disk can remain stably balanced outside the pulsar's light cylinder, in which case excess material might be expelled from the system by the wind from the active pulsar.  Very recently, the detection of X-ray pulsations in J1023 has shown that a trickle of matter is reaching the neutron star surface during the LMXB state \citep{archibald15a}, ruling out models in which none of the accreting material manages to pass the centrifugal barrier.   After submission of this manuscript, coherent X-ray pulsations from XSS J12270 were also found \citep{papitto15a}, highlighting the remarkable similarity between these two systems.

Radio imaging observations provide a complementary viewpoint on the accretion processes in LMXBs, since the radio emission is thought to be primarily generated in outflowing material.  The most luminous (as measured from their X-ray emission) neutron star LMXBs, known as the ``Z-sources'', have all shown radio continuum emission, which in a number of cases has been resolved into collimated jets \citep[e.g.][]{fender98a,fomalont01a,spencer13a}.  Several of the more numerous, lower-luminosity ``atoll sources'' have also shown continuum radio emission, although other than a marginal detection in Aql X-1 \citep{miller-jones10a}, these have not been spatially resolved, and their intrinsic radio faintness has typically precluded detailed study.  Black hole LMXBs show significantly brighter radio continuum emission than their neutron star counterparts \citep{migliari06a}, which at high luminosities has been spatially resolved into extended jets \citep{dhawan00a,stirling01a}.  At lower luminosities, the continuum radio emission from black hole systems is attributed to conical, partially self-absorbed jets \citep{blandford79a}, owing to the flat radio spectra, the high radio brightness temperatures, and the unbroken correlation between radio and X-ray emission \citep[$L_{\rm r}\propto L_{\rm X}^{0.7}$; e.g.][]{corbel00a,gallo03a}.  While this correlation is seen to hold down to the lowest detectable X-ray luminosities in the black hole systems \citep{gallo06a,gallo14a}, the radio behaviour of neutron stars has not to date been studied at X-ray luminosities $\lesssim10^{35}$\,erg\,s$^{-1}$.  The solid surface of a neutron star should make the accretion flow radiatively efficient, in contrast to the radiatively inefficient accretion flows around quiescent black holes, and simple accretion theory then predicts a steeper radio/X-ray correlation index for neutron star systems, of $L_{\rm r}\propto L_{\rm X}^{1.4}$ \citep[e.g.][]{migliari06a}.  While possible correlations between the radio and X-ray emission have been reported for two systems \citep{migliari03a,tudose09a}, they were only derived over a small range in X-ray luminosity, and differ markedly in the reported correlation indices.  Deep radio imaging observations of low-luminosity neutron star systems would ascertain whether they are also able to launch jets when accreting far below the Eddington luminosity, and via comparisons with the behaviour of their black hole counterparts, could determine the role played by the event horizon, the depth of the gravitational potential well, and the stellar magnetic field, in launching jets.

A study of J1023 has the potential to answer several of the outstanding questions concerning pulsar recycling and neutron star accretion physics posed above.  The system consists of PSR~J1023+0038, a ``fully" recycled MSP (where fully recycled is usually defined as $P\lesssim10$ ms) with $P=1.69$ ms, plus a stellar companion with a mass of 0.24\Msun\ \citep{archibald09a,deller12b}. Multiple lines of evidence point to the fact that J1023 is a transitional object; the companion appears to be Roche-lobe-filling, and the millisecond pulsar exhibits timing variability consistent with the presence of substantial amounts of ionized material in the system \citep{archibald09a,archibald13a}. Moreover, archival optical and ultraviolet data from the period May 2000 to January 2001 combined with the knowledge that J1023 harbours a neutron star showed that J1023 possessed an accretion disk during this time \citep{thorstensen05a,archibald09a}.  

Recently, the transitional nature was spectacularly confirmed when J1023 abruptly returned to a LMXB-like state, complete with double-peaked emission lines and optical and X-ray flickering indicating an accretion disk, and the disappearance of radio pulsations \citep{stappers14a,halpern13a,patruno14a}.  \citet{archibald15a} show that coherent X-ray pulsations are (intermittently) present in the LMXB state, indicating that material is being accreted onto the neutron star surface, in contrast to most theoretical models.  
These coherent X-ray pulsations are present while X-ray luminosity is a factor of 10--100 lower than that at which pulsations have been seen in other AMXPs \citep[][particularly Figure~5]{archibald15a}. This implies that channeled accretion similar to that seen in higher-luminosity AMXPs is occurring at a much lower accretion rate in J1023 \citep[the amount of material reaching the stellar surface must be in the range $10^{-13} - 10^{-11}$ \Msun\ yr$^{-1}$;][]{archibald15a}, offering the opportunity to develop a more complete understanding of the accretion process across the whole population of LMXBs.  Since J1023 is relatively nearby \citep[$d$ = 1.37 kpc;][]{deller12b}, it is an excellent testbed for studying the accretion process at very low mass-transfer rates, with the fact that the system parameters are precisely known from its time as a radio pulsar \citep{archibald13a} an added bonus. 

The X-ray luminosity (measured in the 1-10 keV band\footnote{Throughout this paper, we use the 1-10 keV band to compare X-ray luminosities between different systems \citep[as used by, e.g.,][]{gallo14a}.}) of J1023 in the LMXB state is $\sim2.3\times10^{33}$ erg/s \citep[][converted from the reported 0.3-10 keV value using webPIMMS\footnote{\url{https://heasarc.gsfc.nasa.gov/cgi-bin/Tools/w3pimms/w3pimms.pl}}]{bogdanov15b},  substantially brighter than the 9$\times10^{31}$ erg/s observed in the radio pulsar state \citep[][converted from the reported 0.5-10 keV value using webPIMMS]{archibald10a}. Optical observations \citep{halpern13a} make it clear that a disk is present, but the details are unclear. Particularly puzzling is a five-fold increase in the system's $\gamma$-ray luminosity \citep[measured in the 0.1--300~GeV range;][]{stappers14a} relative to its radio pulsar state \citep[when the $\gamma$-rays are thought to be produced directly in the pulsar magnetosphere;][]{archibald13a}. No radio pulsations have been detected from the system in the LMXB state \citep{stappers14a}, but even in its radio pulsar state the signal is eclipsed for $\sim$50\% of the orbit at observing frequencies $\lesssim$1.4GHz by ionized material in the system \citep{archibald13a}. Given the clear evidence for accreted material reaching the neutron star surface provided by the X-ray pulsations, it seems inescapable that the radio pulsar mechanism is unable to function, at least a majority of the time, during the LMXB state \citep{archibald15a}.  However, the balance between accreted and ejected material remains unknown, while the possibility remains that the radio pulsar mechanism and associated pulsar wind could reactivate during the brief cessations of X-ray pulsations -- but only if the radio emission is then being scattered or absorbed in intervening ionized material, to remain consistent with the non-detection of radio pulsations.

To investigate the scenarios presented above, radio imaging observations can be employed. A compact jet launched by the accretion disk would exhibit a flat to slightly inverted radio spectrum in the cm range ($0 \lesssim \alpha \lesssim 0.5$, where $\alpha$ is the spectral index and flux density $\propto \nu^\alpha$), due to the superposition of self-absorbed synchrotron components originating in different regions along the jet \citep{blandford79a}.  Depending on the conditions in the jet, optically thin synchrotron emission could also be (intermittently) dominant, with a spectral index $-1 \lesssim \alpha \lesssim -0.5$.  A typical example is the black hole LMXB GX 339--4, whose radio spectral index as monitored over 10 years is consistently bounded by the range $-0.2 \lesssim \alpha \lesssim 0.5$, but showed one excursion into an optically thin regime with $\alpha = -0.5$ \citep{corbel13a}.  If the radio pulsar mechanism remained active, this could be clearly differentiated from jet emission by its much steeper spectrum ($\alpha = -2.8$; \citealp{archibald09a}).  A potential complication is the presence of free-free absorption in intervening material, which could have originated from the companion's Roche-lobe overflow.  This would lead to a suppression of the lower frequency emission, and could make the interpretation of spectra with a limited fractional bandwidth difficult.

In order to determine the presence and nature of any jet launched by accretion onto J1023 and to determine whether the radio pulsar remains (intermittently) active while the disk is present, we made observations with the Karl G. Jansky Very Large Array (VLA), the European VLBI Network (EVN) and the Low Frequency Array (LOFAR) to cover a wide range of timescales, frequencies, and angular resolutions.  We support these radio observations with X-ray monitoring from the \emph{Swift} telescope, and an analysis of the publically available \emph{Fermi} $\gamma$-ray data.  These observations are described in Section~\ref{sec:obs}, and the results are presented in Section~\ref{sec:results}.  Our interpretation of the nature of the radio emission from the system, as well as the implications for our understanding of accreting compact objects generally, are presented in Section~\ref{sec:discussion}.

\section{Observations and data processing}
\label{sec:obs}
We observed J1023 in the radio band a total of 15 times between 2013 November and 2014 April.  These observations are summarized in Table~\ref{tab:observations}.
For all observations, the position for J1023 was taken from the VLBI ephemeris presented in \citet{deller12b}.  During this period, J1023 was also monitored in the 
X-ray band with the {\em Swift} satellite.  The X-ray observations are described in Section~\ref{sec:xray} below.

\begin{deluxetable*}{cccc}
\tabletypesize{\small}
\tablecaption{Radio imaging observations of J1023}
\tablewidth{0pt}
\tablehead{
\colhead{Start MJD} & \colhead{Duration (hours)} & \colhead{Instrument\tablenotemark{A}} & \colhead{Frequency range (GHz)} 
}
\startdata
56606.68	&  2.0	& VLA (B)		& 4.5 -- 5.5, 6.5 -- 7.5  \\
56607.18	&  5.0	& LOFAR		& 0.115 -- 0.162 \\
56607.57	&  1.0	& VLA (B)		& 2.0 -- 4.0 \\
56609.09	&  7.5	& EVN		& 4.93 -- 5.05 \\
56635.52	&  1.0	& VLA (B)		& 8.0 -- 12.0 \\
56650.39	&  1.0	& VLA (B)		& 8.0 -- 12.0 \\
56664.31	&  1.0	& VLA (B)		& 8.0 -- 12.0 \\
56674.40	&  1.0	& VLA (B)		& 8.0 -- 12.0 \\
56679.46 	&  2.5	& VLA (BnA)	& 1.0 -- 2.0, 4.0 -- 8.0, 12.0 -- 18.0  \\
56688.38	&  1.0	& VLA (BnA)	& 8.0 -- 12.0 \\
56701.21	&  1.0	& VLA (BnA)\tablenotemark{B}	& 8.0 -- 12.0 \\
56723.35	&  1.0	& VLA (A)		& 8.0 -- 12.0 \\
56735.28	&  1.0	& VLA (A)		& 8.0 -- 12.0 \\
56748.06	&  1.0	& VLA (A)		& 8.0 -- 12.0 \\
56775.20	&  1.0	& VLA (A)		& 8.0 -- 12.0 
\enddata
\tablenotetext{A}{VLA = The Karl G. Jansky Very Large Array, configuration indicated in parentheses; EVN = The European VLBI Network; LOFAR = The Low Frequency Array.}
\tablenotetext{B}{Observation made during the move from BnA to A array, and hence had reduced sensitivity and sub-optimal $uv$ coverage.}
\label{tab:observations}
\end{deluxetable*}

\subsection{VLA observations}
Our VLA observations were made using Director's Discretionary Time (DDT) under the project codes 13B-439 and 13B-445.  All VLA observations made use of the source J1024--0052 as a gain calibrator,  with a calibration cycle time ranging from 10 minutes at the lower frequencies to 2.5 minutes at 12--18 GHz.  Pointing solutions were performed every hour during the observation on MJD 56679 (which included 12--18 GHz scans).  The source 1331+305 (3C286) was used as both bandpass and primary flux reference calibrator at all frequencies.  Additionally, from the 3rd VLA epoch onwards, a scan on J1407+2827 was included to calibrate the instrumental polarization leakage. The VLA observations were reduced in CASA \citep{mcmullin07a}, using a modified version of the standard VLA pipeline\footnote{https://science.nrao.edu/facilities/vla/data-processing/pipeline} which incorporated additional steps for calibration of instrumental polarization, following standard procedures and using J1331+3030 as the polarization angle calibrator.  After an initial pipeline run, the data were inspected and additional manual editing was performed to remove visibilities corrupted by radio frequency interference, after which the basic calibration was re-derived.  For each frequency, a concatenated dataset including all data at that band was produced in order to make the most accurate possible model of the field.

\begin{figure*}
\includegraphics[width=1.0\textwidth]{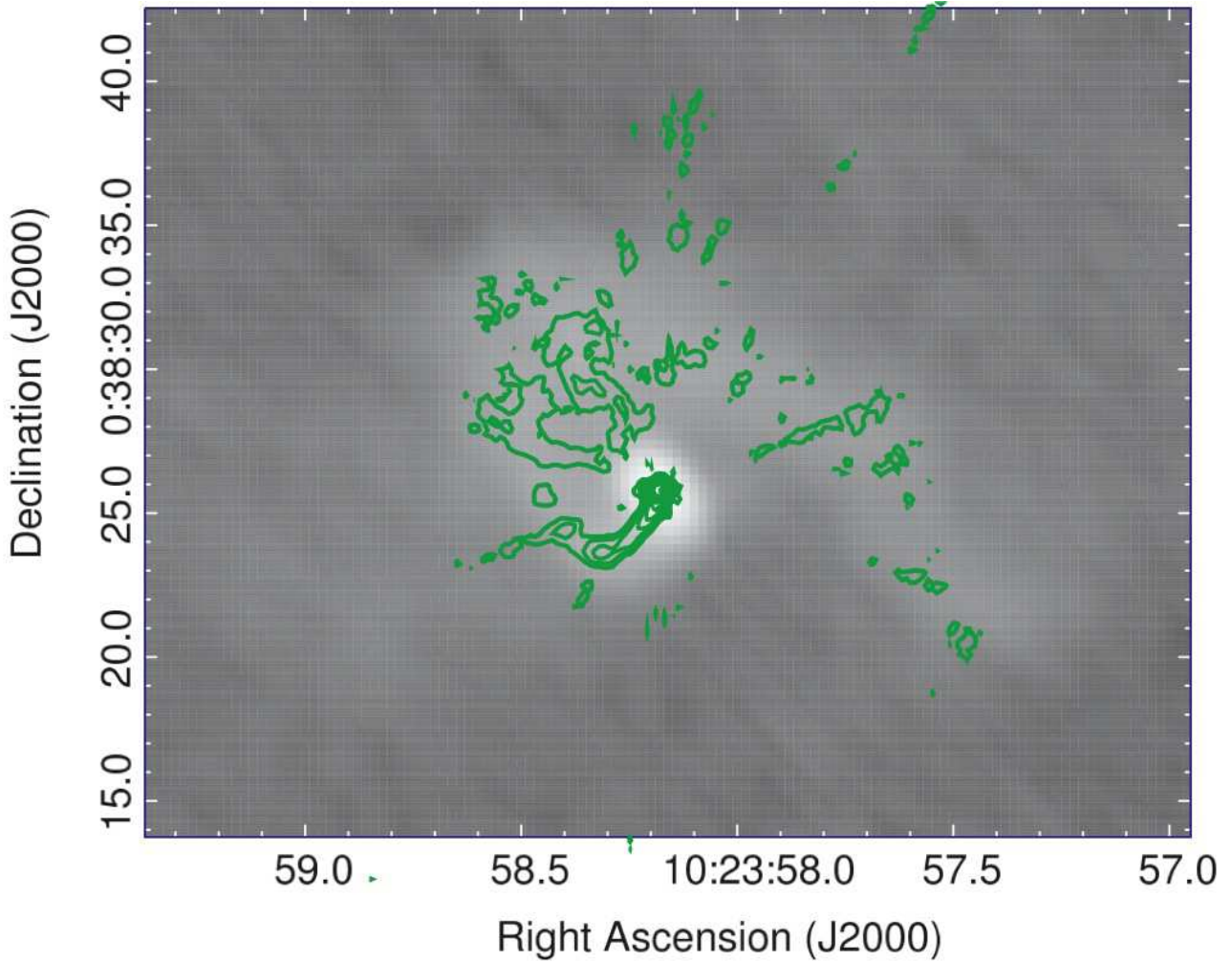} 
\caption{VLA image of the source J102358.2+003826, which dominates the field at frequencies above 2 GHz.  The greyscale shows 4-8 GHz emission on a logarithmic scale, while the contours show the emission at 8-12 GHz.  The first contour is at 3$\sigma$ (8 $\mu$Jy beam$^{-1}$) and the contours increase by factors of 2.  Much of the extended emission is lost in the 8-12 GHz image, due to a combination of higher resolution and the steep spectrum ($\alpha \lesssim -1$) of the extended emission.  An accurate model of this source was necessary to avoid sidelobe contamination at the position of J1023.
}
\label{fig:inbeam}
\end{figure*}

The brightest source in the field surrounding J1023 is J102358.2+003826, a galaxy at redshift 0.449 \citep[obtained with SDSS spectroscopy;][]{ahn14a} with a flux density of 30 mJy at 5 GHz and a steep spectral index ($\alpha = -$0.8). At frequencies above 2 GHz it dominates the field model we construct for self-calibration. It is resolved by the VLA at all frequencies and in all configurations, and the use of a multi-frequency, multi-scale clean (nterms = 2 in the clean method of CASA) was essential to obtain an acceptable Stokes I model with the wide bandwidths available to the VLA.  This model reveals that the bright core is less steep ($\alpha \sim$ $-$0.4) than the surrounding arcsecond-scale extended emission ($\alpha \lesssim -$1.0), as shown in Figure~\ref{fig:inbeam}.  An iterative procedure of self-calibration and imaging was used at all bands.    Depending on the frequency (and hence field of view), a number of other background sources are visible; the source J102348.2+004017  (intrinsic flux density $\sim$200 $\mu$Jy at 10 GHz, intrinsic spectral index $-$0.7) is separated from J1023 by 2 arcminutes and is visible in all epochs.  We refer to this source hereafter as the ``check" source; the fitted brightness of this source (which should not vary in time during an observation) is used to ensure that our absolute flux calibration is adequate.

\begin{figure}
\includegraphics[width=0.48\textwidth]{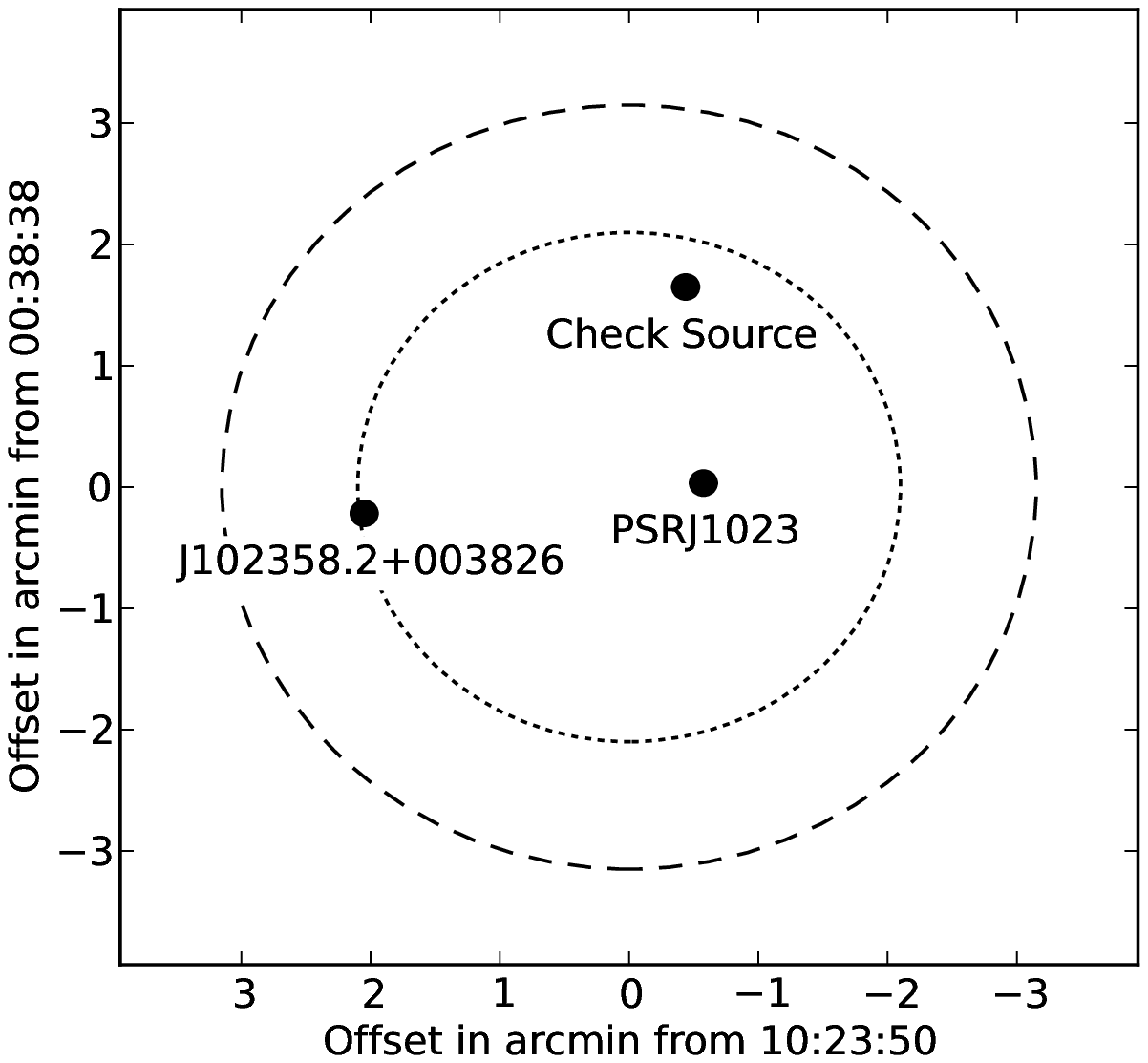} 
\caption{Location of the target source, the brightest source in the field (J102358.2+003826) and the ``check" source, for the 8--12 GHz VLA observations.  
The pointing center was slightly offset from J1023 to ensure that the response in the direction of J102358.2+003826 was not
too greatly attenuated, whilst maintaining near-optimal sensitivity at the position of J1023.  The dashed and dotted lines show the half power
point of the VLA primary beam at 8 GHz and 12 GHz respectively.  Both J102358.2+003826 and the check source sit close to the edge of the primary beam, and hence
their measured flux density can be substantially affected by pointing errors (see text for details).  
}
\label{fig:pointinglayout}
\end{figure}

With the Stokes I field model and calibrated datasets in hand, the following self-calibration and imaging procedure was followed for each epoch:
\begin{enumerate}
\item The data were self-calibrated and all sources except J1023 and the check source were subtracted.  At frequencies of 8 GHz and above, a solution interval of 75 seconds for phase calibration and 45 minutes for amplitude + phase calibration was used; at 4 -- 8 GHz, the phase self-calibration interval was 30 seconds and the amplitude self-calibration interval was 10 minutes, and below 4 GHz the self-calibration interval was 15 seconds and the amplitude self-calibration interval was 5 minutes.
\item In the 8-12 GHz epochs, the amplitude self-calibration was inverted and the inverted solutions applied to the subtracted dataset, to undo the effects of the amplitude self-calibration.  This was necessary due to the location of the source J102358.2+003826 at or beyond the half-power point of the antenna primary beam, meaning that pointing errors dominated the amplitude self-calibration solutions.  Application of these solutions to the target or check source direction would be counter-productive, so they were only used to obtain an accurate subtraction of J102358.2+003826.  At lower frequencies, J102358.2+003826 is well inside the primary beam and this inversion is unnecessary.
\item Multi-frequency clean with nterms=2 was used to derive the average brightness and spectral index of J1023 and the check source over the entire observation in Stokes I.  When cleaning, a clean box with the same size as the synthesized beam was placed on the known source position.  The reference frequency was set to 1.5 GHz, 3 GHz, 6 GHz, 10 GHz or 15 GHz, depending on the recorded band.
\item If J1023 was not detected, an estimate of the noise $\sigma$ was made using a 60x60 pixel area in the center of the residual image, and a 3$\sigma$ upper limit was recorded. 
\item If J1023 was detected, then a point source model was fit to the image using the task imfit and the flux density (taken from the fitted peak brightness, since the source is unresolved), spectral index, and spectral index error were recorded.  This step was also always performed for the check source.
\item For the epochs when J1023 was brightest, we reimaged the data in Stokes I, Q, U and V, using nterms=1 since multi-frequency deconvolution with multiple Taylor terms cannot be performed on multiple Stokes parameters.
\item Additionally, a light curve was made in Stokes I by cleaning short segments of time.  The granularity of the light curve depended on the flux density during the epoch, and ranged from 30 seconds during the brightest epochs to $\sim$ 8 minutes (the scan duration) during the faintest epochs. As will be shown below, the spectral index of J1023 was typically very flat, and accordingly this step was performed with a simple clean (nterms=1).
\item For both J1023 and the check source, the flux density was estimated for each time slice using the task imfit as in point 5 above.  The error in the flux density was estimated from the rms in an off-source box, as the brightness error reported by imfit is typically underestimated unless a large region is included in the fit (in which case the convergence of the fit was often unreliable).
\item The measured flux densities and spectral indices were corrected for the effects of the primary beam, using a simple Airy disk model of the beam which assumed an effective diameter of 25m.
\end{enumerate}

Using the check source, it was possible to examine the consistency of both the overall amplitude scale between epochs and the amplitude scale within an epoch.  For every epoch, we compared the fitted flux density of the check source (with errors) at each time slice to the average value over the whole epoch, and computed the reduced $\chi^2$ assuming a constant value.  The reduced $\chi^2$ obtained in this manner ranged between 0.7 and 1.6 for the 10 GHz observations, consistent with a constant flux density to within the measurement error given the small sample size of 10-35 measurements per epoch.  Accordingly, the amplitude scale within a single observation at 10 GHz is reliable.  For the single observation on MJD56606 at 5 and 7 GHz, the reduced chi--squared of the check source amplitude is 2.1, which may be indicative of an incomplete model for J102348.2+004017 at this frequency, and means that the flux density errors in this observation are possibly slightly underestimated.  However, this does not affect the conclusions that follow.

Between epochs, however, the measured flux density of the check source at 10 GHz varied between 90 and 160 $\mu$Jy, and is inconsistent with a constant value, indicating an epoch-dependent absolute flux calibration error of up to $\sim$25\% in several epochs (although 8 of the 10 epochs fall within the range 110 - 130 $\mu$Jy).  This can be readily understood by inspecting Figure~\ref{fig:pointinglayout}, which shows that the check source is close to the half-power point of the antenna primary beam, and so pointing errors of tens of arcseconds (typical for the VLA without pointing calibration) will lead to considerable amplitude variations (25\% for a pointing error of 30\arcsec).  However, these pointing errors would not lead to similarly dramatic amplitude calibration uncertainties at the position of J1023, since it is much closer to the pointing center (up to a maximum of 10\% for a pointing error of 30\arcsec).  We estimate that the absolute calibration of the J1023 flux values varies betweeen epochs by 5-10\%, and include a 10\% error contribution in quadrature when quoting average flux density values for an epoch.

\subsection{EVN observations}
J1023 was observed with the very long baseline interferometry
technique (VLBI), with the European VLBI Network (EVN) in
real-time e-VLBI mode on 2013 November 13 between 2:00--10:00 UT.
The following telescopes of the e-EVN array participated:
Jodrell Bank (MkII), Effelsberg, Hartebeesthoek, Medicina, Noto,
Onsala (25m), Shanghai (25m), Toru\'n, Yebes, and the phased-array
Westerbork Synthesis Radio Telescope (WSRT). The total data rate
per telescope was 1024~Mbit/s, resulting in 128~MHz total bandwidth
in both left- and right-hand circular polarization spanning 4.93 -- 5.05 GHz, using 2-bit
sampling. Medicina and Shanghai produced the same bandwidth using
1-bit sampling and a total data rate of 512~Mbit/s per telescope.
The target was phase-referenced to the nearby calibrator J1023+0024
in cycles of 1.5 min (calibrator) -- 3.5 min (target). The position of
J1023+0024 was derived from the astrometric solution of \citet{deller12b}. Check scans
on the calibrator J1015+0109, with coordinates taken from the NASA Goddard Space Flight
Center (GSFC) 2011A astro solution, were also included.

The data were correlated with the
EVN Software Correlator \citep[SFXC;][]{pidopryhora09a} at the
Joint Institute for VLBI in Europe (JIVE) in Dwingeloo, the Netherlands.
The data were analysed in the 31DEC11 version of AIPS \citep{greisen03a} using the
ParselTongue adaptation of the EVN data calibration pipeline
\citep{reynolds02a, kettenis06a}.  Imaging was performed in version 2.4e of Difmap \citep{shepherd94a} using 
natural weighting.  J1023 was tentatively detected with a peak brightness of
50$\pm13$~$\mu$Jy beam$^{-1}$ (3.8$\sigma$ significance) at the expected
position (which is known to an accuracy much better than the synthesized beam size \citep{deller12b}, making
a 3.8$\sigma$ result significant).

To assess the quality of phase-referencing in our VLBI experiment, we
compared the phase-referenced and phase self-calibrated e-EVN data on
J1015+0109. This revealed that the amplitude losses due to residual
errors in phase-referencing did not exceed the $10\%$ level. The difference
between the phase-referenced and the a-priori assumed coordinates were
0.25~mas in right ascension and 2.5~mas in declination; the latter
slightly exceeds the 3$\sigma$ error quoted in the GSFC astro solution,
and it is probably because of the difference between the true and the
assumed tropospheric zenith delay during the correlation. This source
position shift is nevertheless within the naturally weighted beamsize
of 6.1$\times$3.1~mas, major axis position angle 81~degrees. 
Finally, in addition to the e-EVN data, we also processed the
WSRT local interferometer data in AIPS. Comparison of WSRT
and e-EVN flux densities of compact calibrators showed that the
absolute calibration of the VLBI amplitudes were accurate to $\sim4\%$.

\subsection{LOFAR observations}
\label{sec:lofar}
J1023 was observed by LOFAR \citep{van-haarlem13a} for 5 hours on 2013 November 13,
between 04:22--09:19 UT (and hence spanning more than one full orbit), in a director's discretionary time observation.  The observing
bandwidth was 48 MHz spanning the frequency range 114-162 MHz.
All Dutch stations were included, for a total of 60 correlated elements.  3C196 was used as a calibrator
source and was observed for 1 minute every 15 minutes.  The field centered on J1023 was iteratively
self-calibrated, beginning with a model built from the revised VLSS catalog \citep{lane14a}, followed
by imaging at low resolution, further self-calibration, and finally imaging at high resolution (using a maximum
baseline length of 25 km).  For the results described below, only the bandwidth range 138.5--161.7 MHz was
used.

When in the radio pulsar state, J1023's rotation-powered radio pulsations were easily detectable using LOFAR's high-time-resolution beam-formed modes 
\citep{stappers11a} and would have been similarly obvious in imaging observations, with a period-averaged 150 MHz
flux density of $\sim$40 mJy (V.~Kondratiev et al., in prep).  In our observations, J1023 was not detected, with 
a 3$\sigma$ upper limit of 5.4 mJy beam$^{-1}$.  
The attained image rms is a factor of 10 higher than the predicted thermal
rms, and is limited by sidelobes from bright sources in the field which are corrupted
by direction-dependent gain errors.  The observation spanned local sunrise, which is a time of rapid 
ionospheric variability and hence worse than usual direction-dependent errors.  
Improvements in LOFAR data reduction techniques since the time of
these observations, including direction-dependent gain calibration,
mean that images which closely approach the thermal noise are now
becoming possible, and we therefore anticipate being able to improve upon 
these limits in a future analysis.

\subsection{\emph{Swift} X-ray observations}
\label{sec:xray}
We analyzed 52 targeted \emph{Swift} observations taken between 2013
October 31 and 2014 June 11, most of which were also presented in
\citet{coti-zelati14a}.
We used all data recorded by the \emph{Swift}/X-ray-Telescope (XRT) 
which always operated in photon-counting mode (PC mode), with a time resolution
of 2.5 s. The cumulative exposure time was $\approx 87$ ks. 
The data were analysed with the {\ttfamily HEASoft}
v.6.14 and {\ttfamily xrtpipeline} and we applied
standard event screening criteria (grade 0--12 for PC data).

We extracted the photon counts from a circular region of size between 20 and 40
arcsec (depending on the source luminosity). The background was calculated by
averaging the counts detected in three circular regions of the same
size scattered across the field of view after verifying that they did
not fall close to bright sources. The 1--10 keV X-ray counts were then
extracted and binned in 10-s intervals with the software {\ttfamily
  xselect}. The X-ray counts were then corrected with the task {\ttfamily
  xrtlccorr}, which accounts for telescope vignetting, point-spread
function corrections and bad pixels/columns and rebinned the counts
in intervals of 50 seconds.

\subsection{\emph{Fermi} $\gamma$-ray monitoring}

We examined the \emph{Fermi} data on J1023 up to 2014 October 20, extending the light curve shown in \citet{stappers14a} to a time baseline in the LMXB state of almost 16 months. As in \citet{stappers14a}, we carried out the analysis in two ways. First we selected a spectral model for J1023 and all sources within 35 degrees. We then divided the time of the \emph{Fermi} mission into segments of length $5\times10^6$ seconds, and using the $0.1$--$300\;\text{GeV}$ photons within 30 degrees of the position of J1023 from each time segment, we fit for the normalizations of all sources within 7 degrees of J1023 (we followed the binned likelihood tutorial on the \emph{Fermi} Cicerone\footnote{\url{http://tinyurl.com/fermibinnedtutorial}}). The normalization of J1023 serves as an estimate of background-subtracted $\gamma$-ray flux. Because this method depends on complex spectral fitting procedures, we estimated the flux using a simple aperture photometry method as well: we selected all photons within one degree of J1023 and $1$--$300\;\textrm{GeV}$ (since photons with lower energies have a localization uncertainty greater than a degree), then used their exposure-corrected flux.

\section{Results}
\label{sec:results}

\subsection{Radio spectral index}
\label{sec:specindex}
\begin{figure}
\begin{center}
\begin{tabular}{c}
\includegraphics[width=0.48\textwidth]{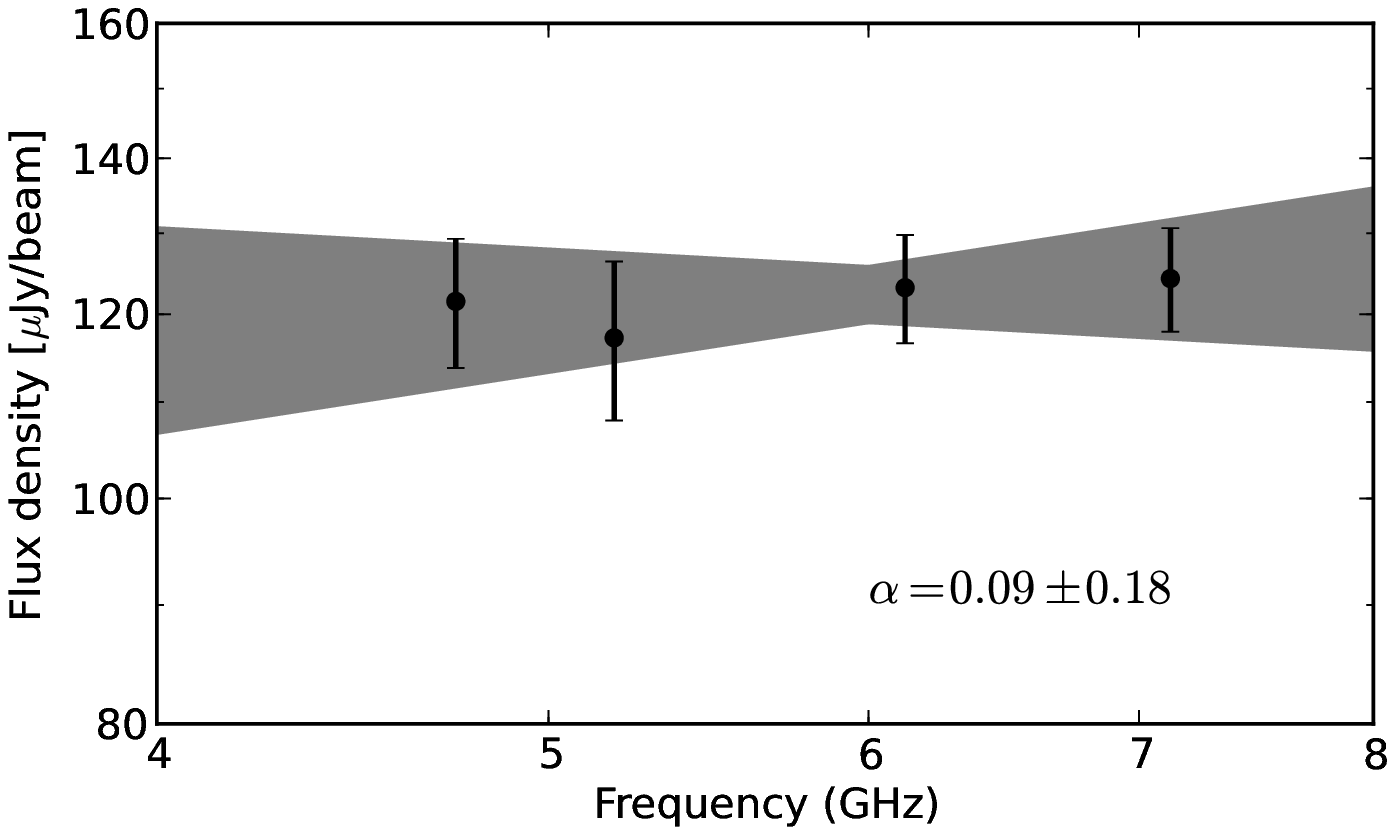} \\
\includegraphics[width=0.48\textwidth]{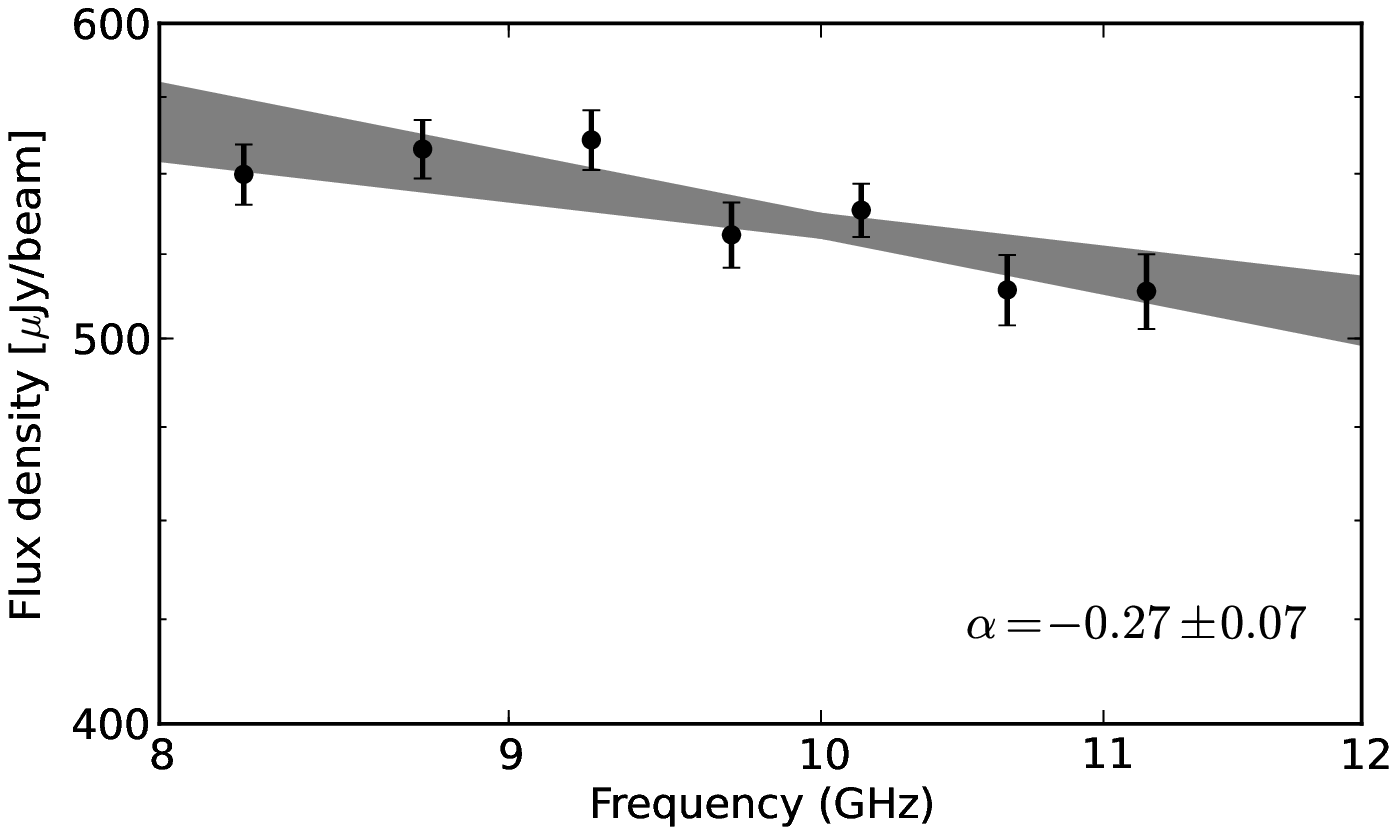} \\
\includegraphics[width=0.48\textwidth]{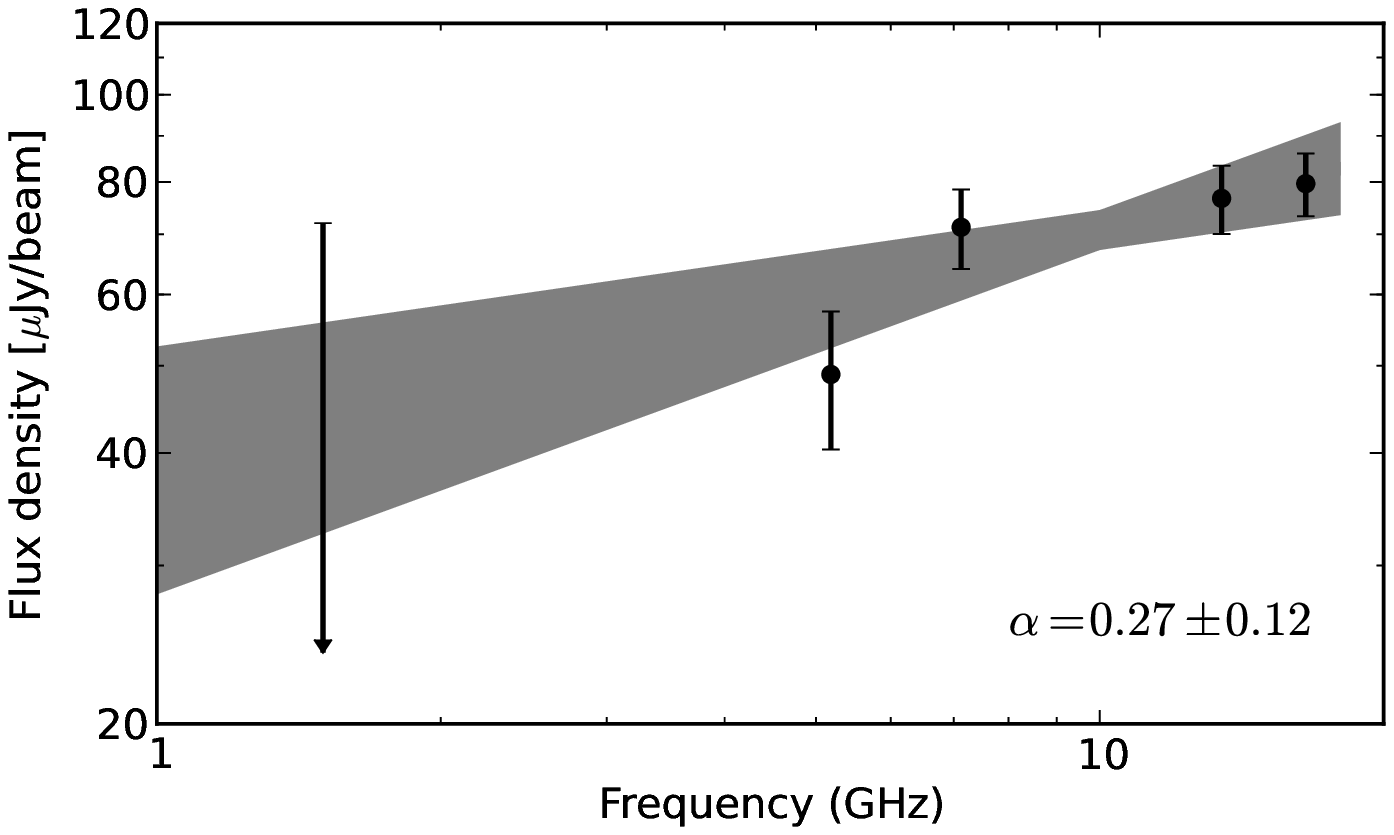}
\end{tabular}
\caption{The radio spectrum of J1023, as seen at 4.5 -- 7.5 GHz during a period of typical flux density (MJD 56606, top),  8 -- 12 GHz during a period of enhanced flux density (MJD 56674, middle), and 1 -- 18 GHz during a period of relatively low activity (MJD 56679, bottom).  The gray shaded region shows the 1$\sigma$ error region (including the fitted errors on reference flux density and spectral index).
}
\label{fig:radiospectrum}
\end{center}
\end{figure}

Figure~\ref{fig:radiospectrum} illustrates the instantaneous radio spectrum obtained for J1023 in 3 different observations: the 4.5 -- 7.5 GHz observation on MJD 56606, the 8-12 GHz observation on MJD 56674 (when the source was brightest), and the 3--frequency observation on MJD 56679.  Table~\ref{tab:averagevals} lists the spectral indices obtained from all of the VLA observations where J1023 was detected.  The median value of $\alpha$ from the 12 observations with a spectral index determination is $0.04$.  The limited precision of the spectral index determination during the fainter epochs makes it difficult to say with certainty within what range the spectral index varies; considering only the reasonably precise measurements (with an error $<$ 0.3) yields a range $-0.3 < \alpha < 0.3$.  In any case, it is certain that there is some variability: the reduced $\chi^2$ obtained when assuming a constant spectral index across all 12 epochs is 3.0, and some of the better-constrained epochs differ at the 3$\sigma$ level.   Within a single epoch, we were able to test for spectral index variability on timescales down to $\sim$5 minutes only for the few observations where J1023 was brightest; here, variation at the 2$\sigma$ level ($\Delta\alpha \simeq \pm0.3$) was seen on timescales of $\sim$30 minutes.  This variability on timescales of minutes to months would be expected if the emission originated in a compact jet, as the observed spectrum is made up of the sum of individual components along the jet (which have different spectra) whose presence and prominence changes with time.

We examined the possibility of a spectral turnover of J1023 at frequencies below 4 GHz, as is expected theoretically for a partially self-absorbed compact jet \citep[e.g.][]{markoff01a}, but not unambiguously detected to date in any compact jets from an LMXB.  Although relatively few deep, low-frequency observations have been made of LMXBs, we note that the bright black hole high mass X-ray binary Cygnus X-1 has been detected at frequencies as low as 350 MHz (G. de Bruyn, priv. comm.)  Only two observations of around 40 minutes on-source each were made of J1023 below 4 GHz, and the sensitivity is somewhat poorer than at the other frequencies due to reduced bandwidth.  In the 3--frequency VLA observation of MJD 56679, the flux density of J1023 was relatively low, and the 1--2 GHz upper limit of 72 $\mu$Jy (3$\sigma$) is consistent with the extrapolated flux density from the higher frequencies (30--50 $\mu$Jy).  In the 2-4 GHz observation on MJD 56607, J1023 was not detected in the average image, with a 3$\sigma$ upper limit of 30 $\mu$Jy beam$^{-1}$.  However, in the corresponding light curve of the 2-4 GHz observations, $\sim$3$\sigma$ peaks at the $\sim$60 $\mu$Jy beam$^{-1}$ level are seen at the position of J1023 in two of the eight scans, and we treat these as possible detections.  In any case, the average flux density during this observation must be considerably lower than the minimum value of $\sim$45 $\mu$Jy seen at higher frequencies in any 40 minute period; if there is no spectral turnover, then J1023 must have been a factor of $\sim$2 fainter in this epoch than in any other observation.  More low-frequency (preferably simultaneous) observations would be needed to definitively determine whether a spectral turnover is present.

\begin{deluxetable*}{cccc}
\tabletypesize{\small}
\tablecaption{Per-epoch average flux density and spectral index values for J1023.  The quoted uncertainties for the average flux density include statistical fit errors and the estimated 10\% uncertainty in the absolute flux density scale; the latter dominates.}
\tablewidth{0pt}
\tablehead{
\colhead{Start MJD} & \colhead{Reference frequency (GHz)} & \colhead{Flux density ($\mu$Jy)} & \colhead{Spectral index} 
}
\startdata
56606.68	&  6	& 122 $\pm$ 12	&  $0.09\pm0.18$ \\
56607.18	&  3	& $<$30 			& -- \\
56635.54 & 10 & 85 $\pm$ 9 & $0.93 \pm 0.48$ \\
56650.41 & 10 & 428 $\pm$ 43 & $-0.18 \pm 0.10$ \\
56664.34 & 10 & 160 $\pm$ 16 & $-0.02 \pm 0.24$ \\
56674.42 & 10 & 533 $\pm$ 53 & $-0.27 \pm 0.07$ \\
56679.50 & 10 & 70 $\pm$ 7 & $0.27 \pm 0.12$ \\
56688.41 & 10 & 101 $\pm$ 10 & $0.15 \pm 0.34$ \\
56701.23 & 10 & 78 $\pm$ 9 & $0.82 \pm 0.64$ \\
56723.37 & 10 & 76 $\pm$ 8 & $0.49 \pm 0.47$ \\
56735.30 & 10 & 100 $\pm$ 10 & $-0.23 \pm 0.35$ \\
56748.09 & 10 & 54 $\pm$ 6 & $-0.31 \pm 0.61$ \\
56775.22 & 10 & 45 $\pm$ 6 & $-0.93 \pm 0.73$
\enddata
\label{tab:averagevals}
\end{deluxetable*}

\subsection{Radio variability}
\begin{figure*}
\begin{center}
\includegraphics[width=0.9\textwidth]{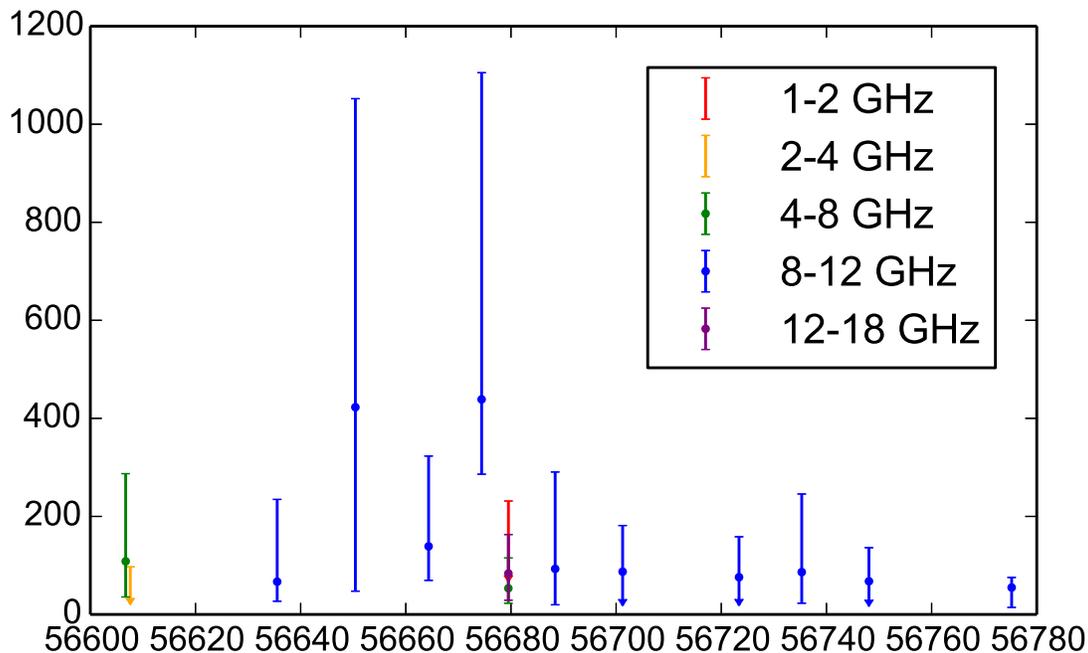} 
\caption{The radio light curve of J1023 from the VLA and EVN observations.  The points show the median values at each epoch, with the error bars encompassing the maximum and minimum values seen within the epoch.  The short timescale variability is within an epoch is shown in Figure~\ref{fig:lightcurvezooms}.
}
\label{fig:lightcurvesummary}
\end{center}
\end{figure*}

\begin{figure*}
\begin{center}
\begin{tabular}{ccc}
\includegraphics[width=0.3\textwidth]{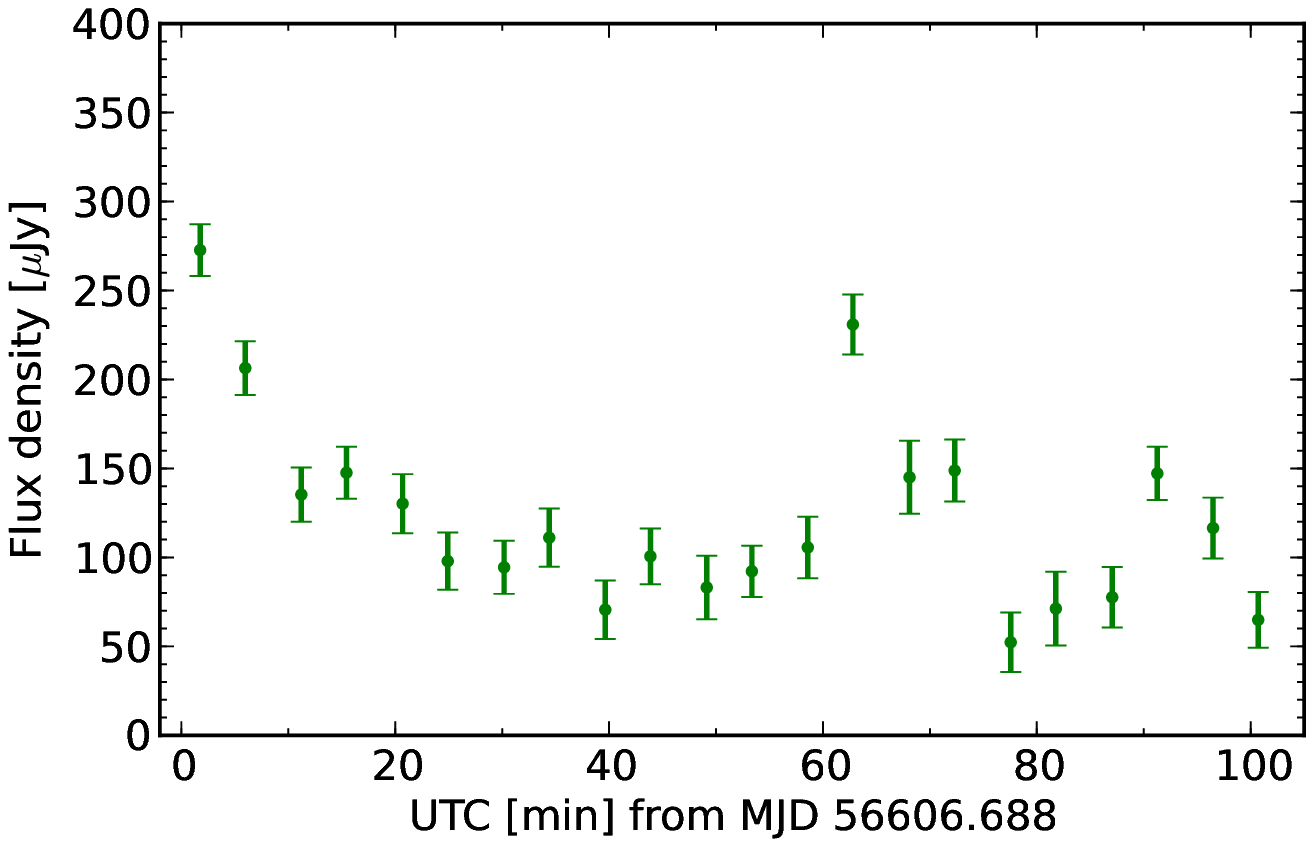} &
\includegraphics[width=0.3\textwidth]{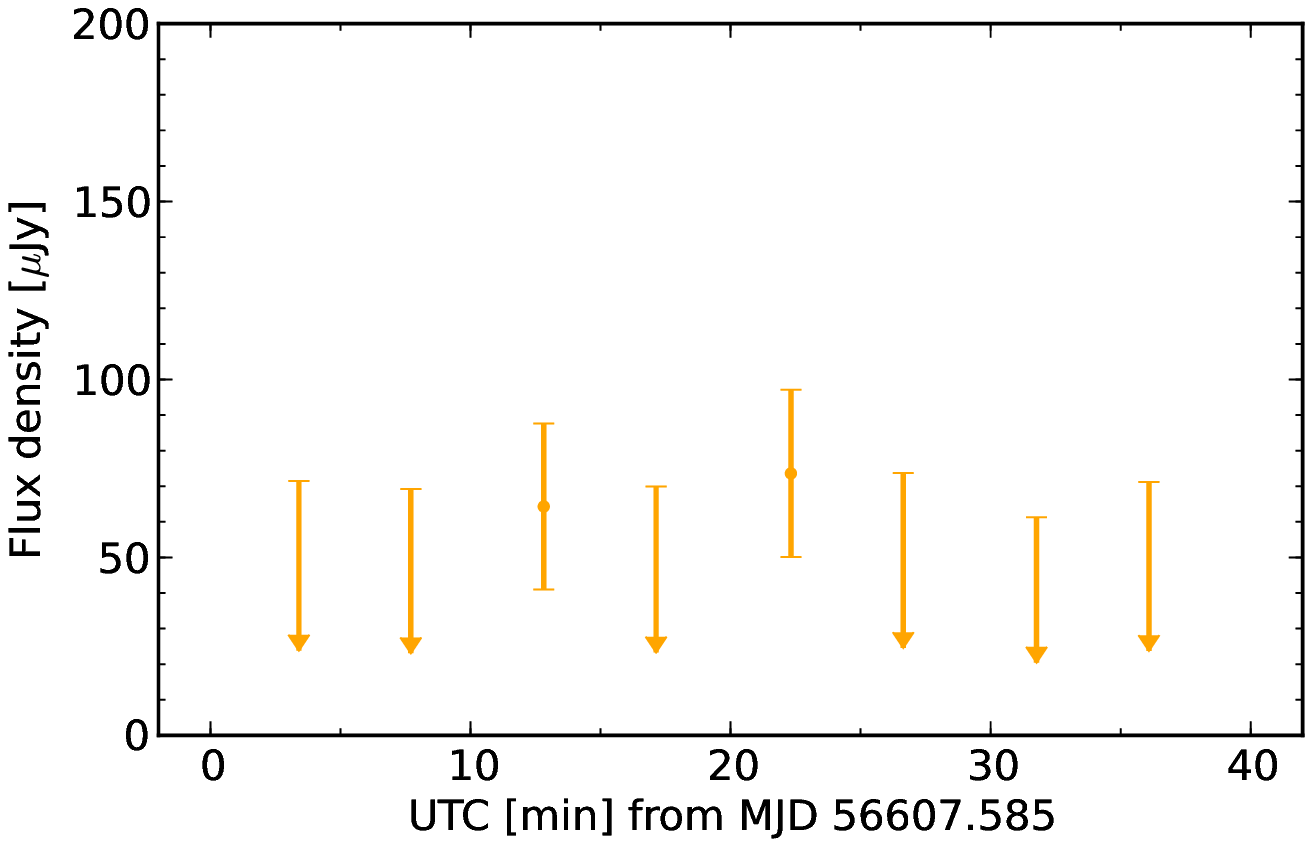} &
\includegraphics[width=0.3\textwidth]{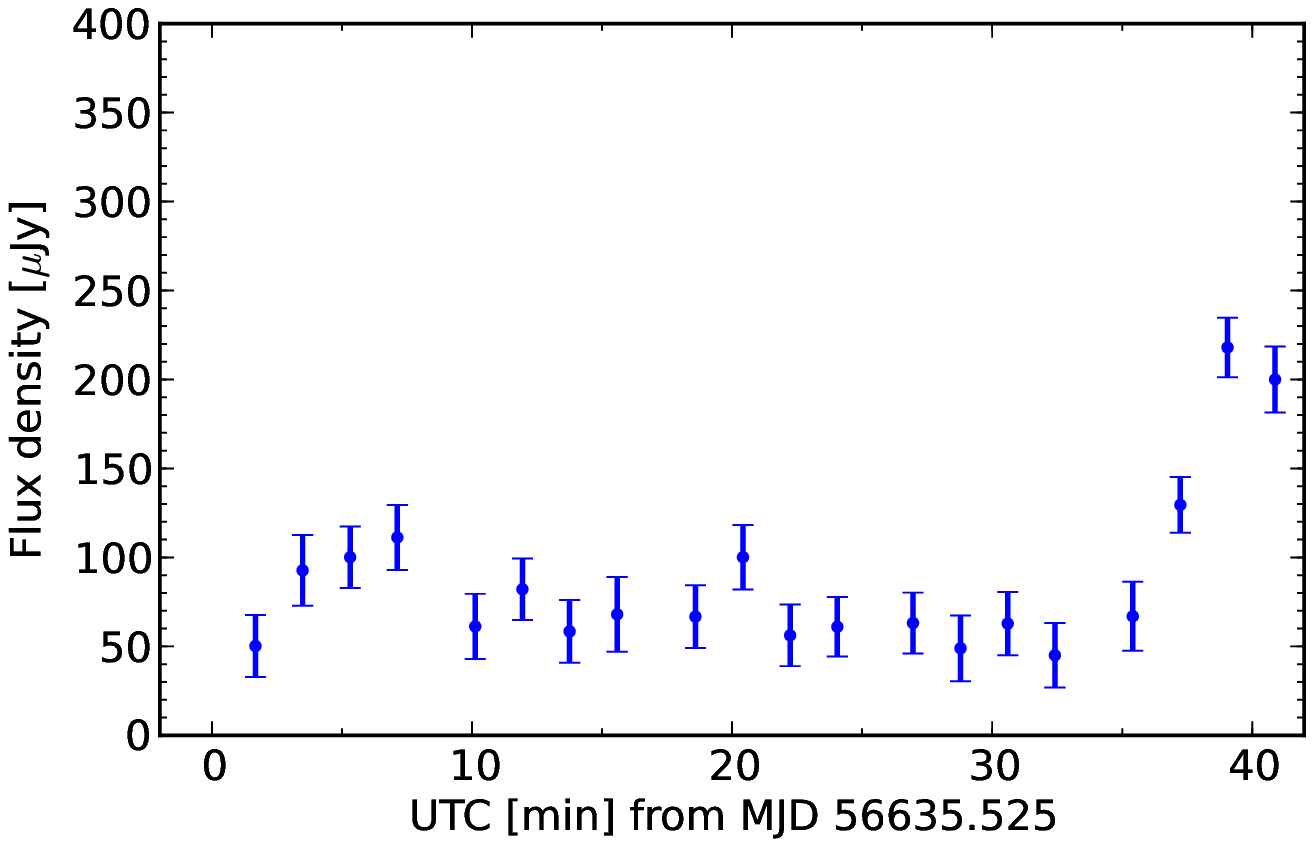} \\
\includegraphics[width=0.3\textwidth]{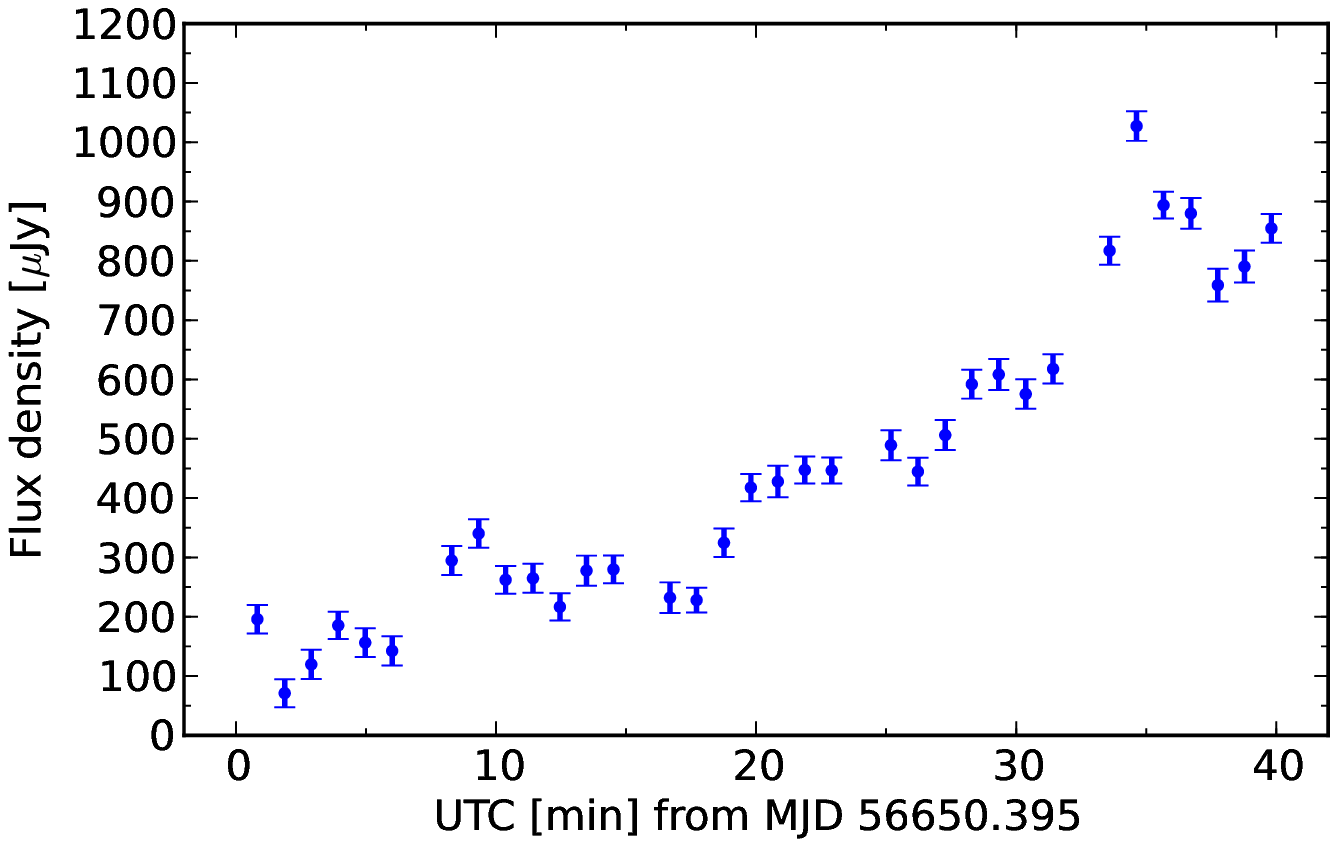} &
\includegraphics[width=0.3\textwidth]{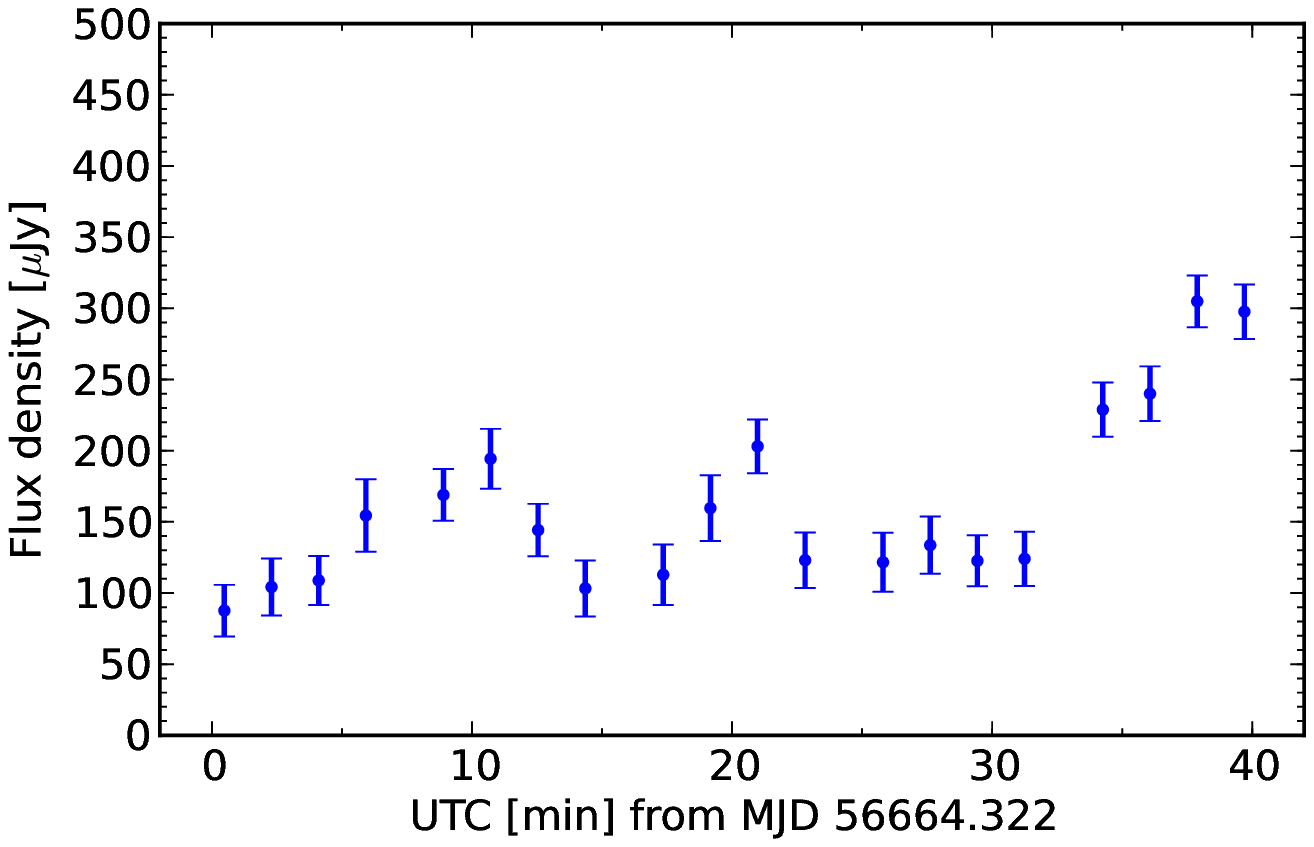} &
\includegraphics[width=0.3\textwidth]{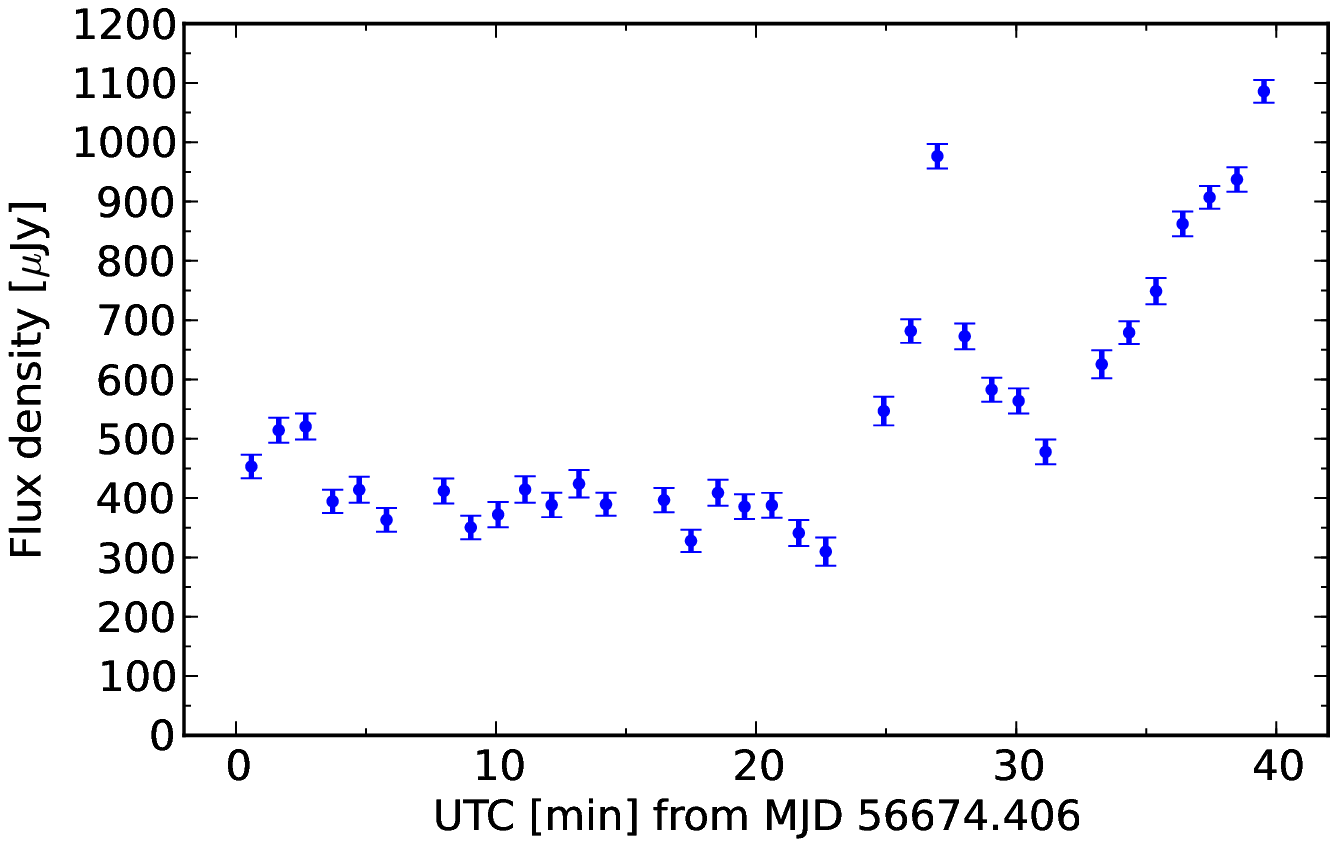} \\
\includegraphics[width=0.3\textwidth]{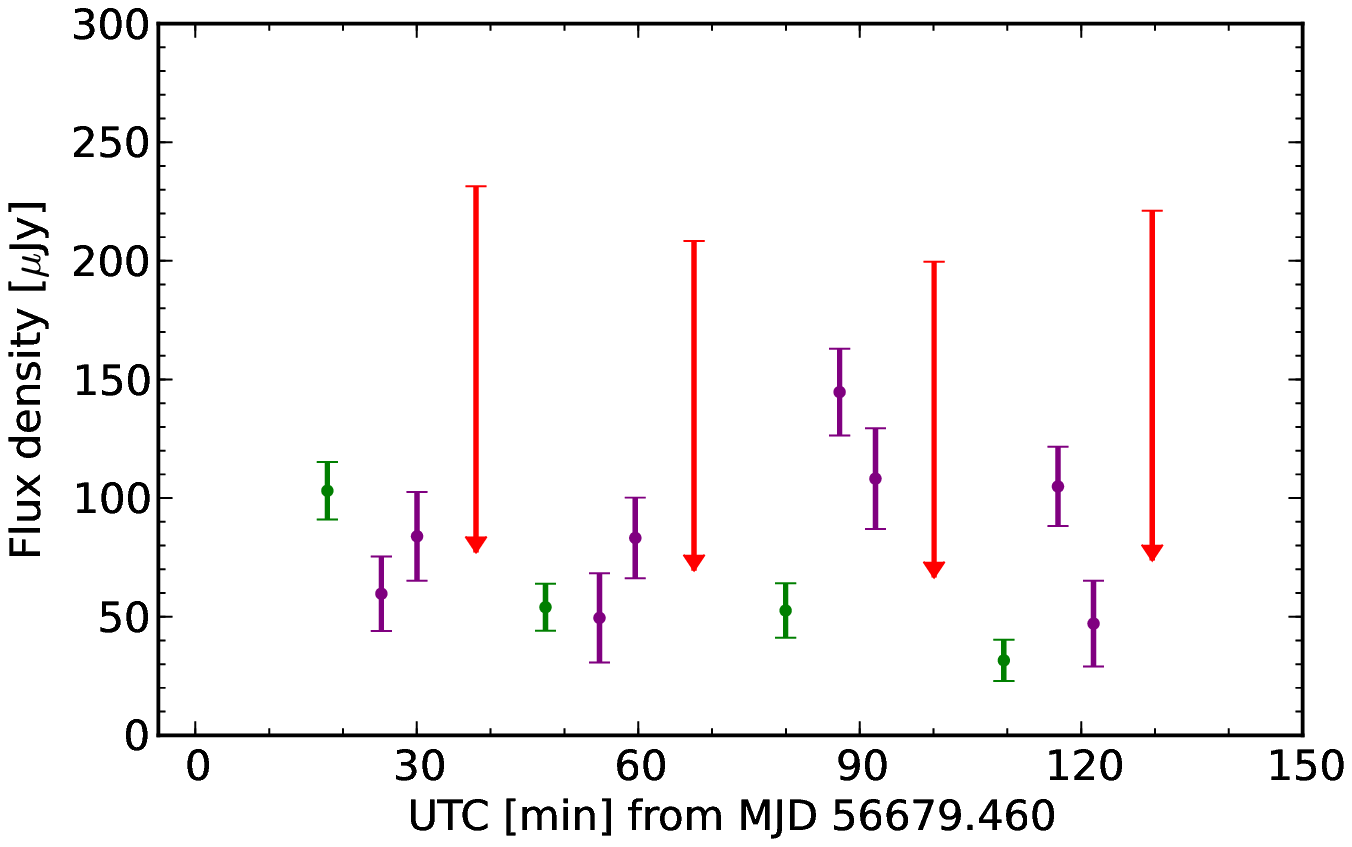} &
\includegraphics[width=0.3\textwidth]{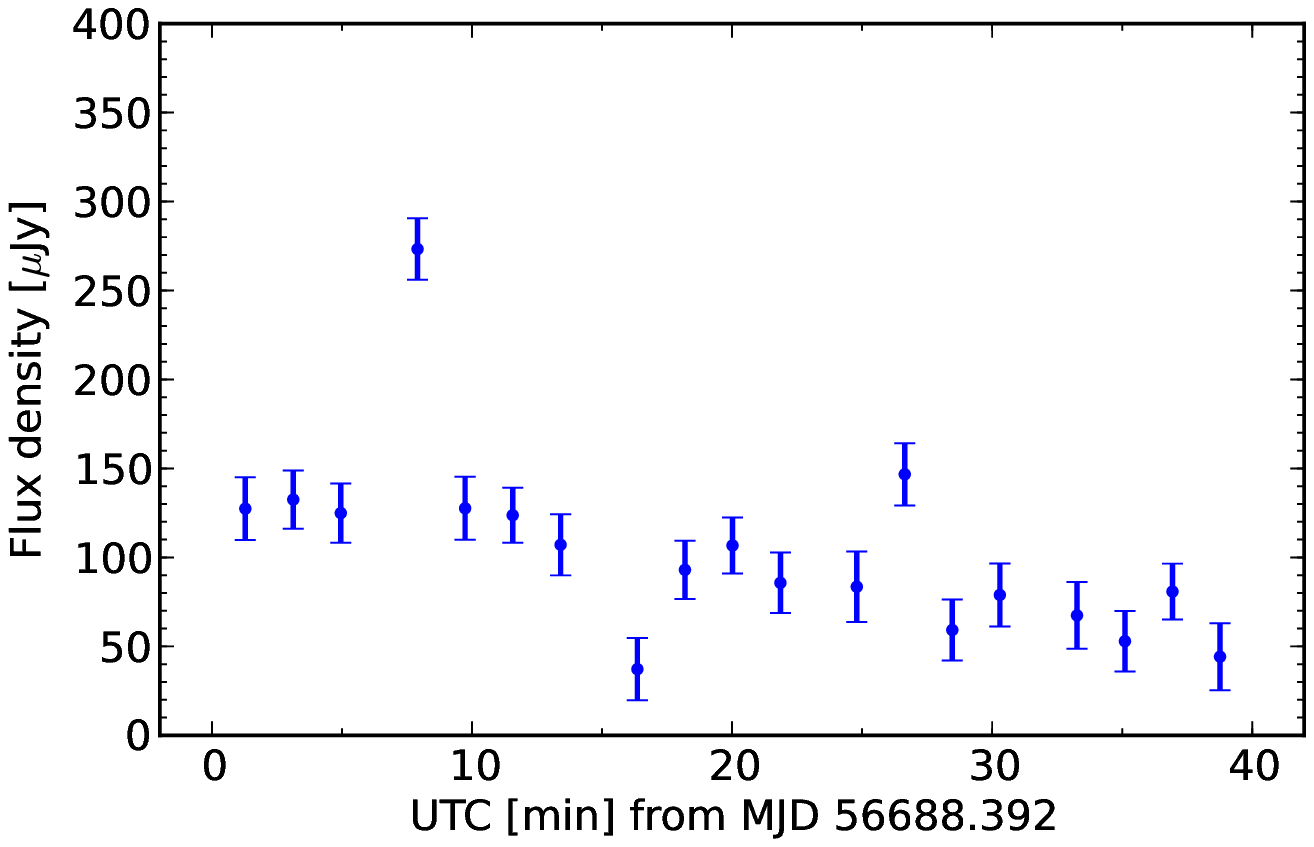} &
\includegraphics[width=0.3\textwidth]{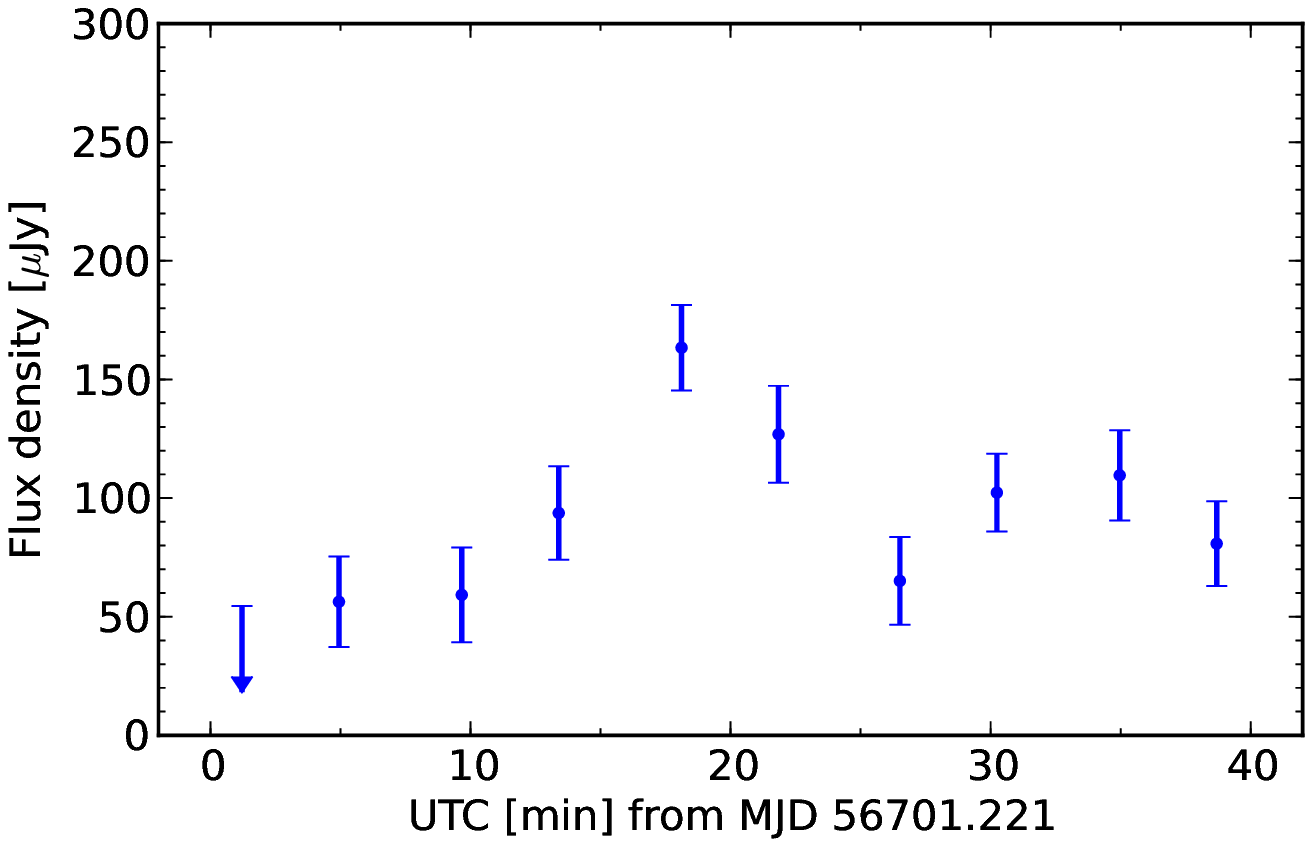} \\
\includegraphics[width=0.3\textwidth]{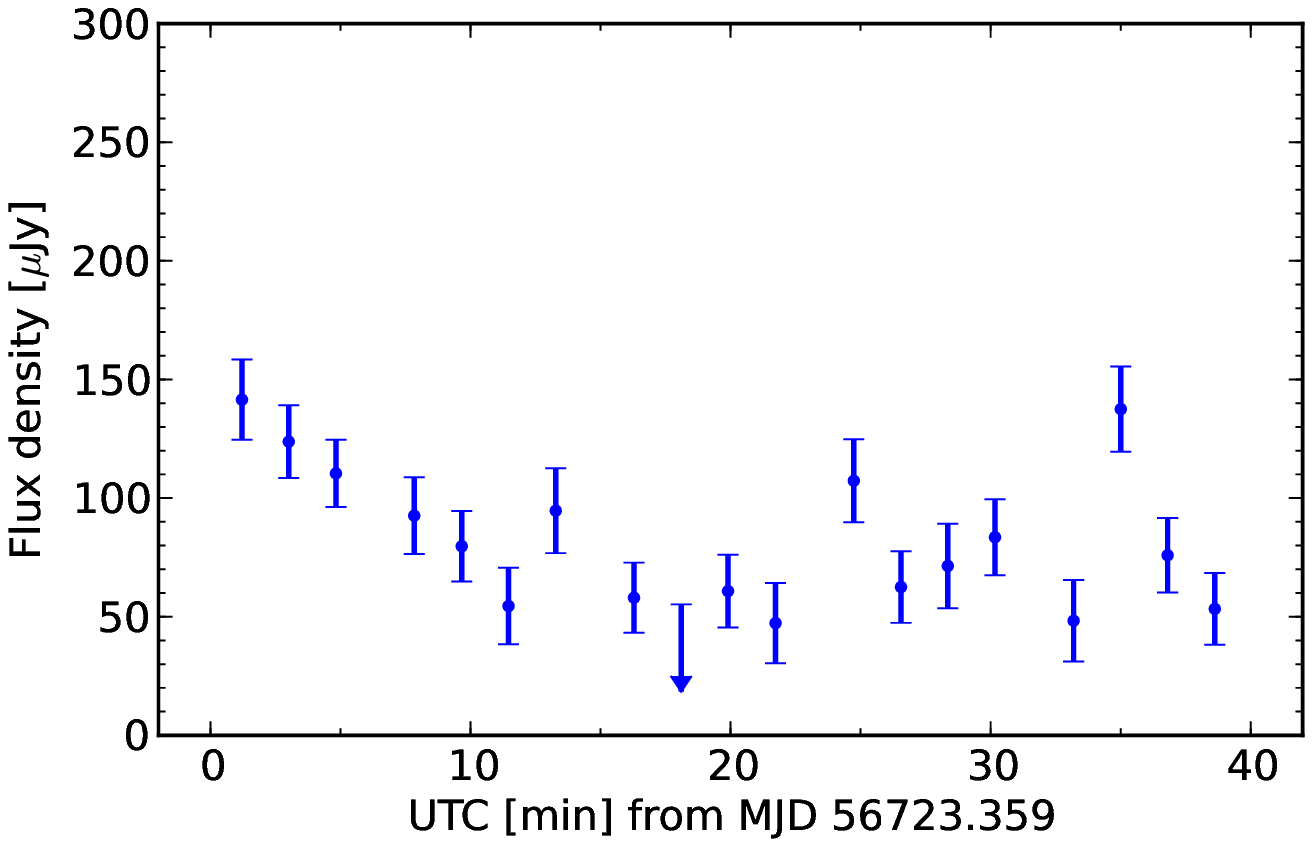} &
\includegraphics[width=0.3\textwidth]{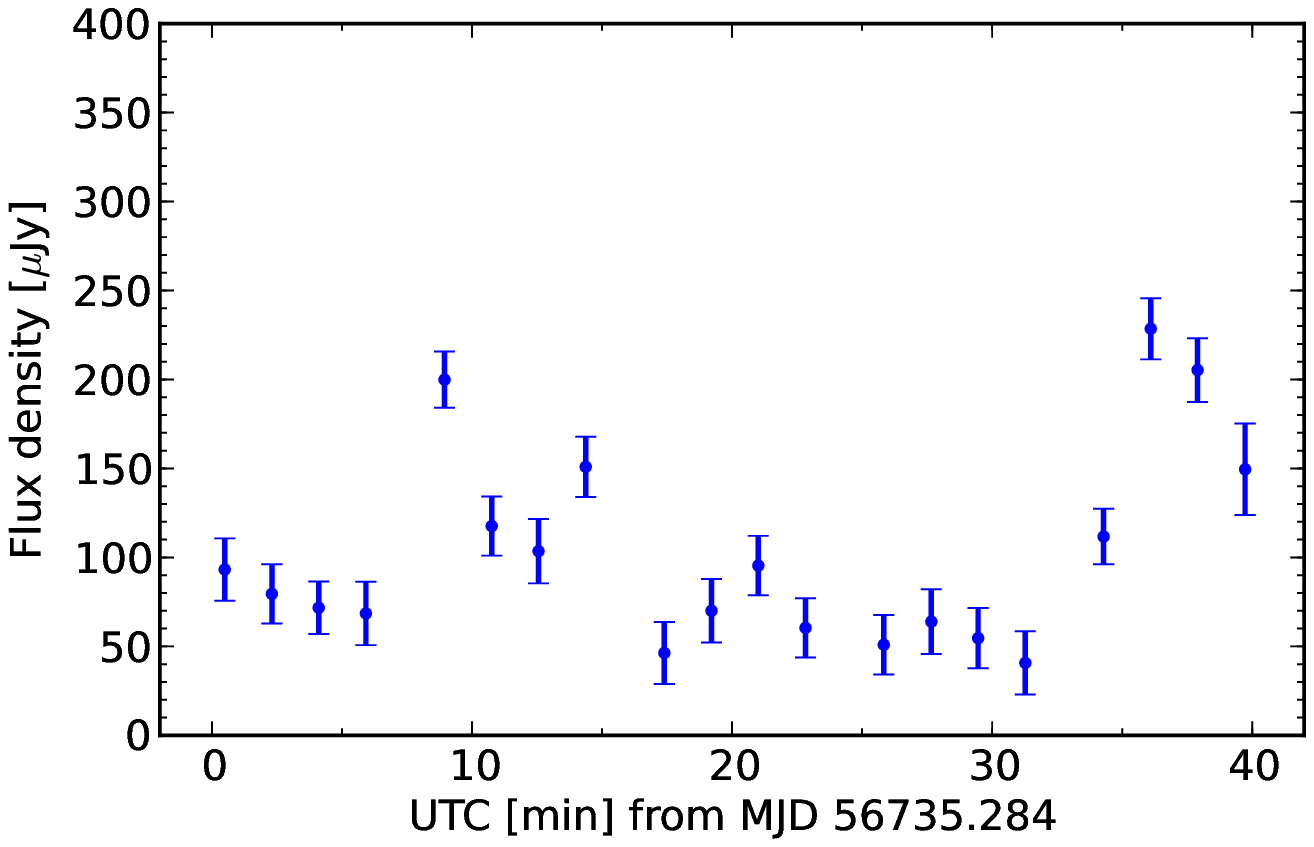} &
\includegraphics[width=0.3\textwidth]{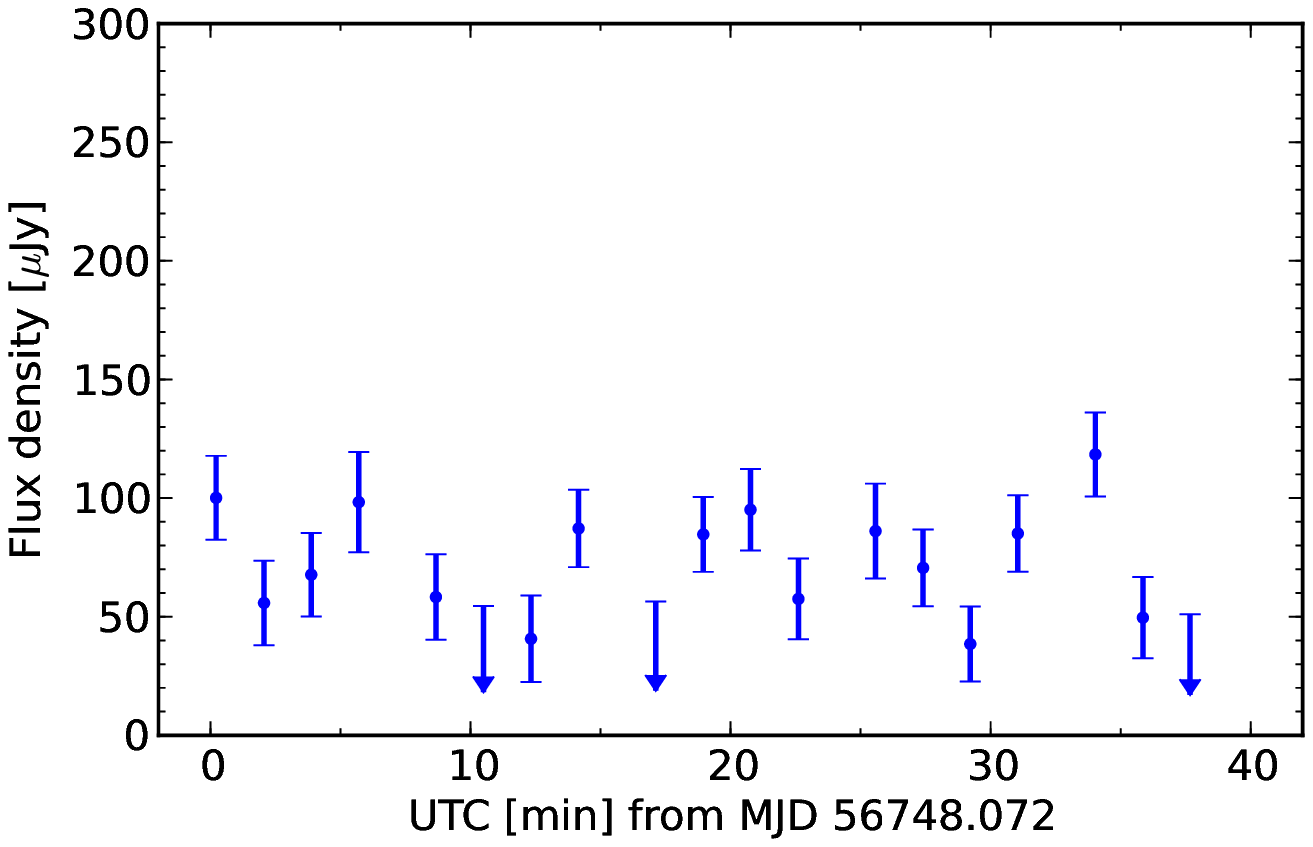} \\
\includegraphics[width=0.3\textwidth]{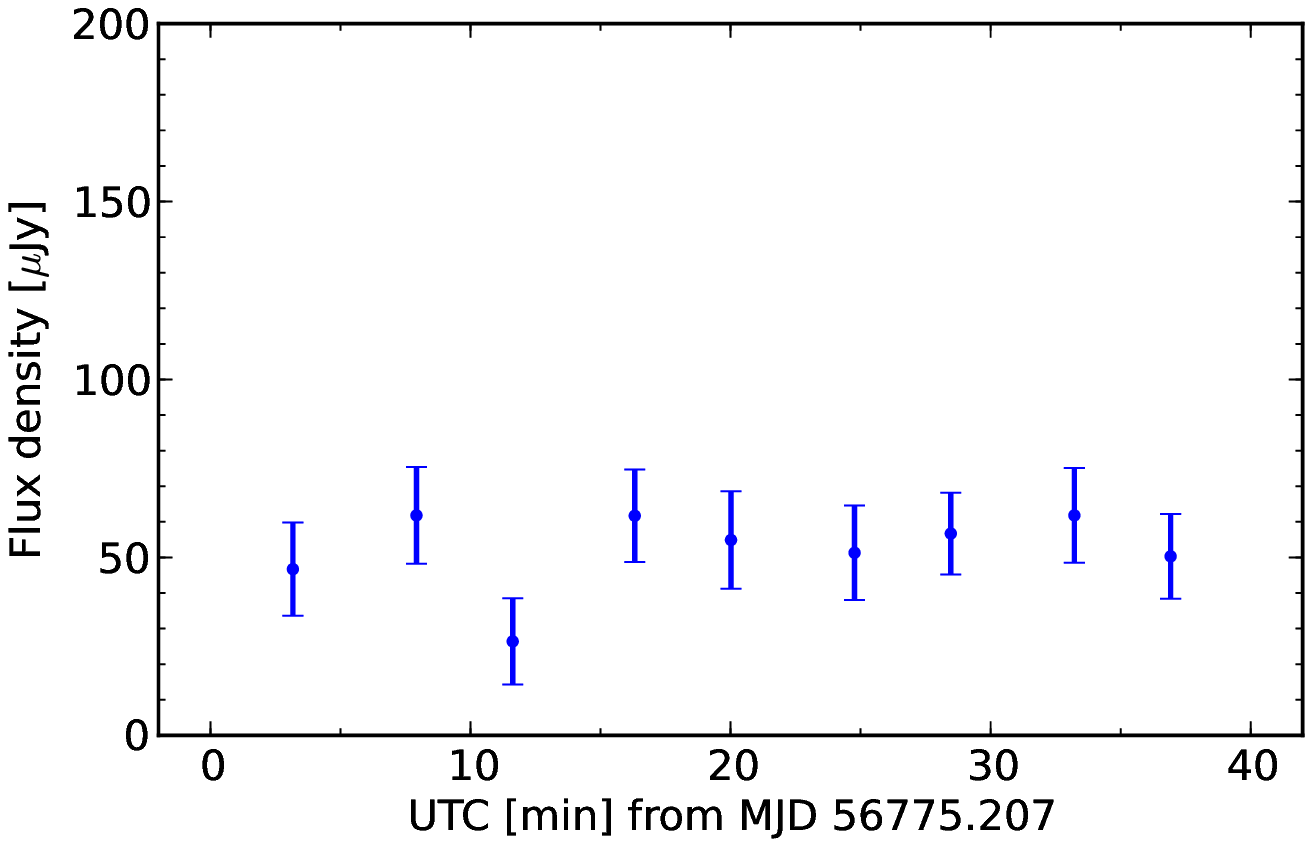} 
\end{tabular}
\caption{The light curves from the 13 VLA observations (where J1023 could be detected on a sub-observation timescale).  The colors representing different frequencies are the same as those used for Figure~\ref{fig:lightcurvesummary}.  Factor-of-two variations are common within minutes, and the flux density increases by an order of magnitude within half an hour in one case (MJD 56650).
}
\label{fig:lightcurvezooms}
\end{center}
\end{figure*}

In the 15 radio observations made in the period 2013 November to 2014 April, the flux density of J1023 varies by almost two orders of magnitude, with factor-of-two variability within 2 minutes and order-of-magnitude variability on timescales of 30 minutes.  Figures~\ref{fig:lightcurvesummary} and~\ref{fig:lightcurvezooms} show the radio lightcurve over this six month period; note the rapid variability and high flux density during the observation on MJD 56674.  From light travel time arguments, we can therefore constrain the maximum source size to be 120 light-seconds, $\sim$30 times the binary separation of 4.3 light-seconds \citep{archibald13a}, although it could of course be considerably smaller. Given the observed flux density of 1 mJy, the brightness temperature must reach at least $3\times10^{8}$ K.  If we neglect the effects of relativistic beaming \citep[the system inclination is known to be 42 degrees;][]{archibald13a}, then under the typical assumption of a maximum brightness temperature of $10^{12}$\,K (as expected for unbeamed, incoherent, steady-state synchrotron emission) we can calculate the minimum source size to be $\sim$2 light-seconds: around half the binary separation, and 7500 times larger than the pulsar light cyclinder \citep[81 km;][]{bogdanov15b}.

We also searched for a frequency-dependent time delay in our VLA light curves.  For the observation on MJD 56674 where the source was brightest and shows the most rapid variability, we imaged the data from 8--10 GHz and 10--12 GHz separately with a time resolution of 30 seconds.  We then performed a cross-correlation analysis to determine if any time lag was present at the lower frequency, as might be expected for a jet in which the lower frequency emission primarily originates further from the origin.  No significant offset was detected, which is unsurprising given the 30 second time resolution (the shortest timescale at which we still obtain detections of a reasonable significance) and the source size limits derived above.

The radio flares seen with the VLA (Figure~\ref{fig:lightcurvezooms}) could be related to the observed X-ray flaring activity, but the radio emission from J1023 does not appear to exhibit the short, sharply defined excursions to a lower luminosity mode that are seen in X-ray observations \citep{bogdanov15b}.  Although we lack the sensitivity to probe the radio emission to very faint levels on minute timescales, in the periods of brightest radio emission (MJD 56650 and 56674), we can exclude sharp ``dips" in flux density on timescales of two minutes.  In general, the variation appears to be slower in radio than X-ray, and this slower variation implies that the radio emission region is more extended than the region in which the X-rays are produced.

Table~\ref{tab:averagevals} shows the mean flux density (at the central frequency of 3, 6, or 10~GHz as appropriate) and spectral index observed for J1023 at each epoch.

\subsection{Radio polarization}
Our best constraints on the fractional polarization of the radio continuum emission come from the two epochs when J1023 was brightest (MJD 56650 and 56674).  Although linear polarization was detected from J102358.2+003826 (which, given its location beyond the half-power point of the primary beam, could be instrumental), J1023 showed no evidence for significant linear polarization in either epoch, down to a $3\sigma$ upper limit on the fractional polarization of 3.9 per cent.  To get the best possible limit, we selected only the time periods when the Stokes I emission from J1023 exceeded 500\,$\mu$Jy\,beam$^{-1}$ in each epoch, stacked those data from the two epochs, and made a combined image in Stokes I, Q, U and V.  This allowed us to reduce our $3\sigma$ limit on the fractional polarization to 24\,$\mu$Jy\,beam$^{-1}$ (3.4 per cent).  While the data from MJD 56674 showed possible evidence for Stokes V emission at 23\,$\mu$Jy\,beam$^{-1}$ (4.1 per cent), the check source showed a consistent level of Stokes V emission at that epoch (9.1 $\mu$Jy\,beam$^{-1} = 4.6$ per cent), strongly suggesting that this could be caused by errors in the gain calibration or instrumental leakage correction.

Linear polarization has been detected in compact jets from hard-state black hole X-ray binaries \citep[e.g.][]{corbel00a,russell14a}, at levels of up to a few per cent.  Therefore, while our limits on the polarization of J1023 are consistent with the levels expected from a partially self-absorbed compact jet, they can neither definitively confirm nor rule out this scenario.

\subsection{X-ray variability}

\begin{figure*}
\begin{center}
\includegraphics[width=0.9\textwidth]{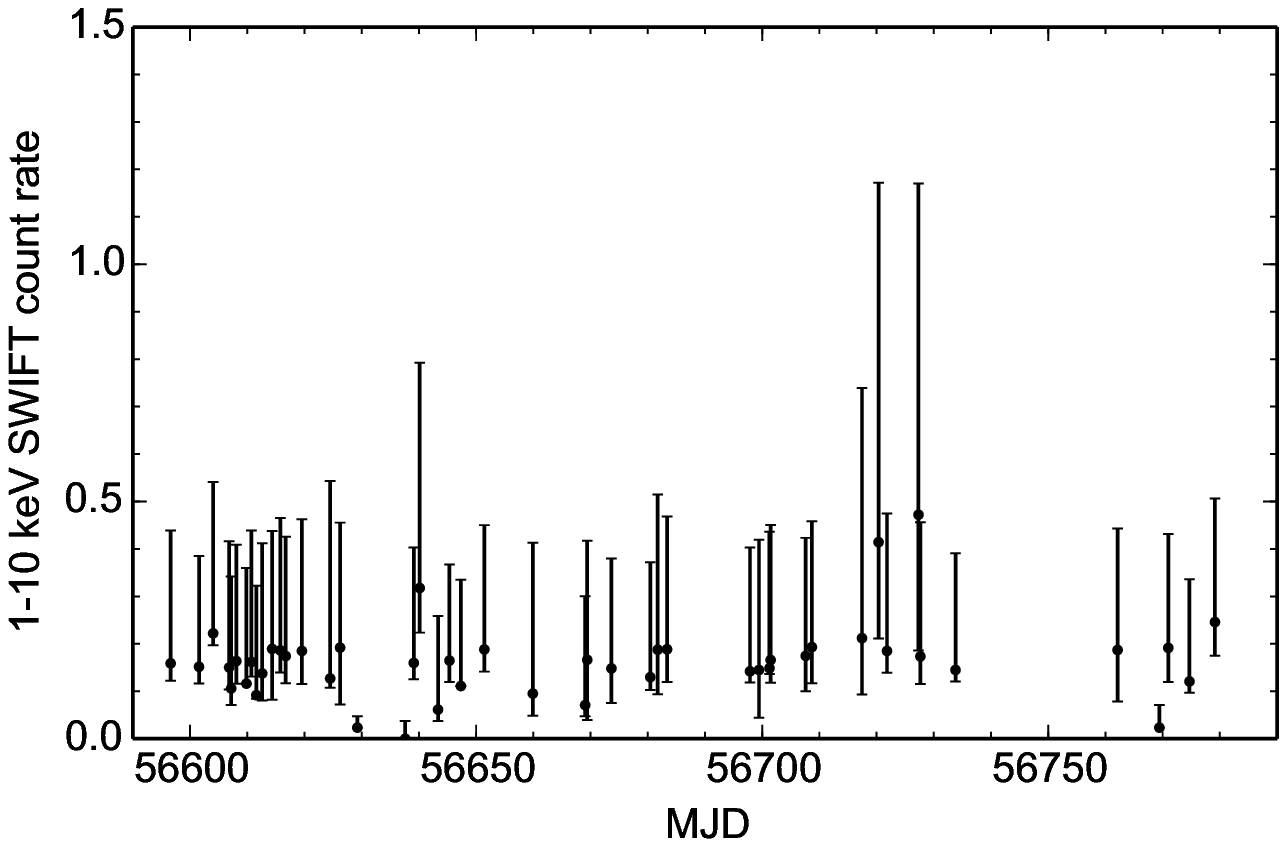} 
\caption{The X-ray light curve of J1023.  The points show the median value at each epoch, with the error bars showing the range encompassing 67\% of the points in the observation.  Examples of the short-term variability within an observation are shown in Figure~\ref{fig:xraylightcurvezooms}.
}
\label{fig:xraylightcurvesummary}
\end{center}
\end{figure*}

\begin{figure*}
\begin{center}
\begin{tabular}{cc}
\includegraphics[width=0.48\textwidth]{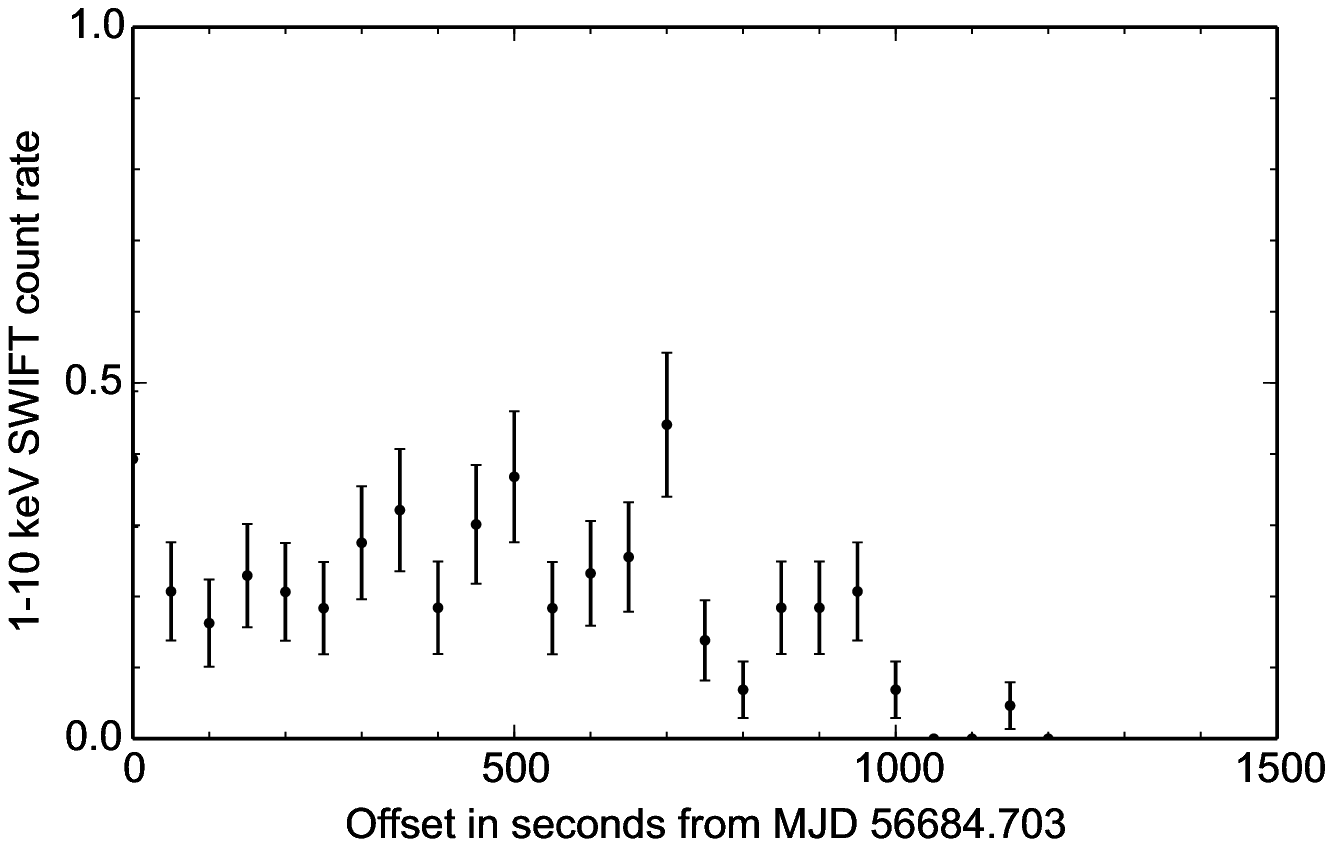} & 
\includegraphics[width=0.48\textwidth]{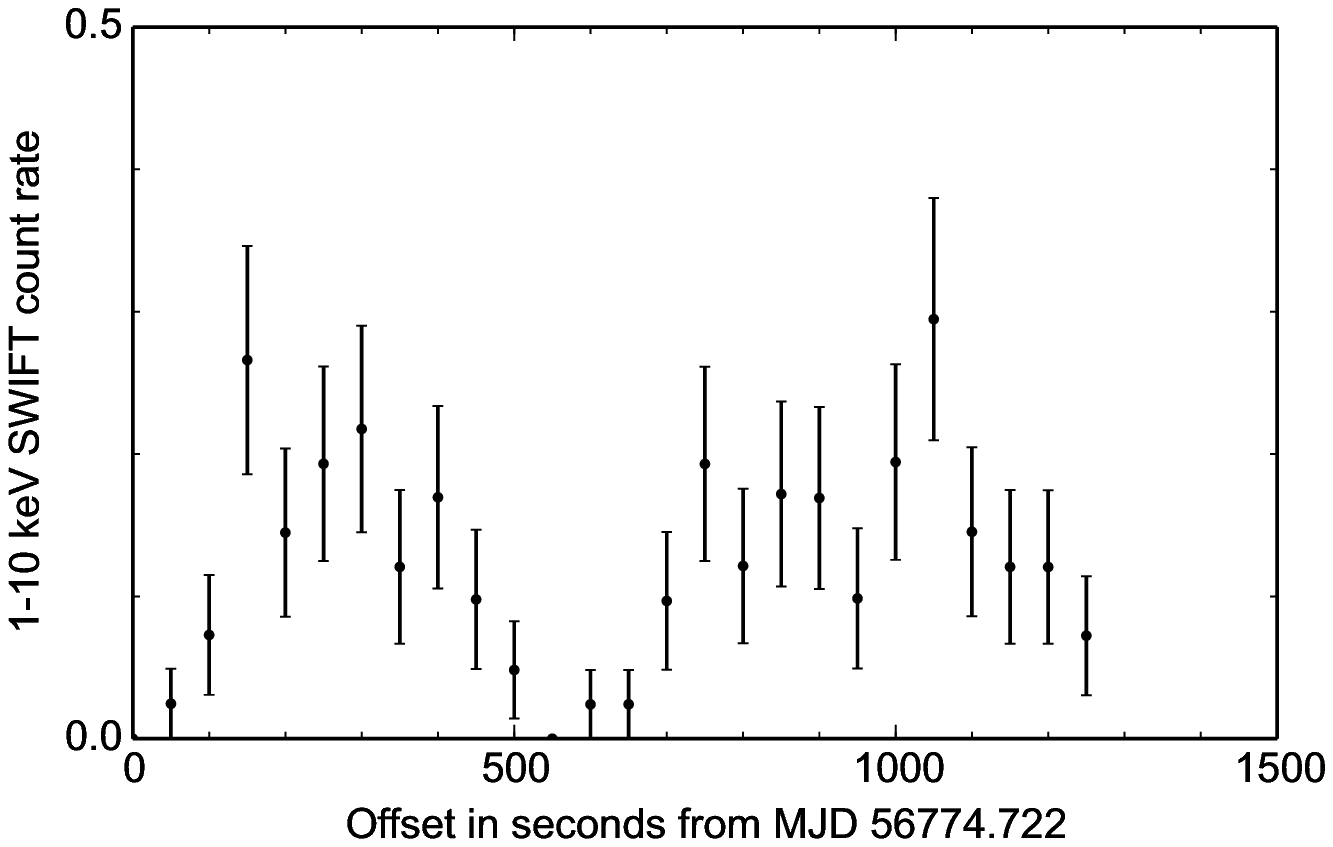} 
\end{tabular}
\caption{The light curves from two arbitrarily selected \emph{Swift} observations, showing the variation in count rate on minute timescales.}
\label{fig:xraylightcurvezooms}
\end{center}
\end{figure*}

Figure~\ref{fig:xraylightcurvesummary} shows the X-ray light curve of J1023 as observed by \emph{Swift}.  Figure~\ref{fig:xraylightcurvezooms} shows two epochs, each of duration $\sim$20 minutes, to highlight the variability seen on short timescales.  We note that higher sensitivity X-ray observations of J1023 made with \emph{XMM-Newton} reveal much shorter-timescale structure in the X-ray lightcurve, showing that J1023 spends most of its time in a stable ``high" mode, with brief drops down to a lower-luminosity ``low" mode and infrequent higher-luminosity ``flares" \citep{bogdanov15b}.  The switches between the modes occur on a timescale of tens of seconds, and the duration of the modes typically ranges from tens of seconds to tens of minutes.
This trimodal X-ray behavior seen in J1023 \citep[and also, very recently, in XSS J12270;][]{papitto15a} differs markedly from the behavior of black hole LMXBs, where variability of comparable magnitude and timescale has been seen in quiescence but without the quantization into distinct modes \citep[e.g., V404 Cygni,][]{bradley07a}.  While some other black hole LMXBs such as A0620--00 and V4641 Sagitarii have shown distinct optical modes which indicate changing accretion properties \citep[e.g.,][]{cantrell08a,macdonald14a}, the timescales over which these persist are far longer -- weeks to months.
  
We converted the \emph{Swift} XRT count rate to 1--10 keV luminosity using webPIMMS and a power-law model with $N_\mathrm{H} = 5\times10^{20}$ cm$^{-2}$ and photon index $\Gamma = 1.56$, as obtained by an analysis of {\em Swift} data on J1023 by \citet{coti-zelati14a}. The median luminosity seen in the {\em Swift} observations is $2.0\times10^{33}$ erg s$^{-1}$, in agreement with the ``high" mode luminosity of $2.3\times10^{33}$ erg s$^{-1}$ measured by \citet{bogdanov15b} and the mean luminosity of $2.4\times10^{33}$ erg s$^{-1}$ measured by \citet{coti-zelati14a}, where the luminosity was converted to 1-10 keV in both cases.

\subsection{\emph{Fermi} $\gamma$-ray increase}
\begin{figure*}
\begin{center}
\includegraphics[width=0.9\textwidth]{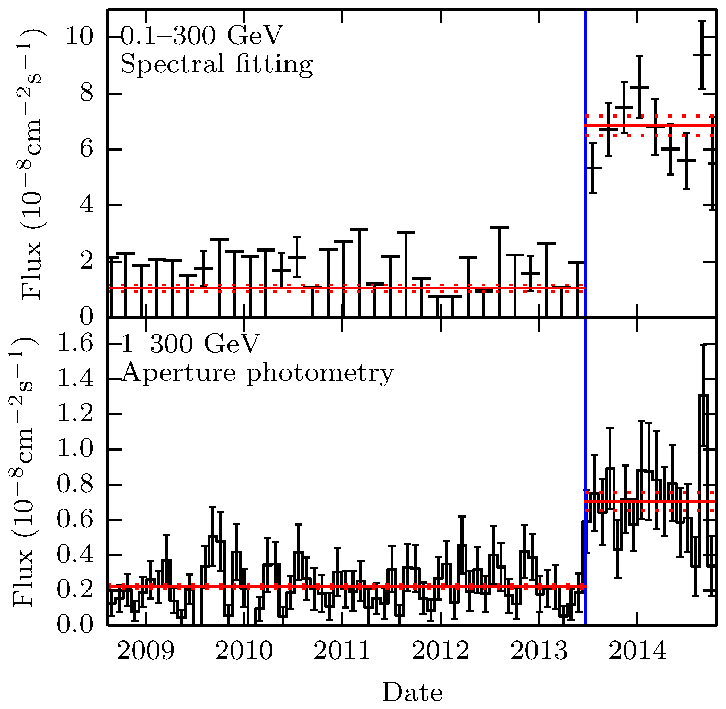} 
\caption{The \emph{Fermi} $\gamma$-ray light curve of J1023, extended from that presented in \citet{stappers14a}.  Top panel: $\gamma$-ray photon fluxes obtained by fitting for the normalization of a spectral model. Bottom panel: $\gamma$-ray photon fluxes obtained by measuring numbers of counts within a one-degree aperture; this panel considers only photons of $>1\,\textrm{GeV}$ since lower-energy photons are too poorly localized. The vertical line marks the disappearance of radio pulsations from J1023 in 2013 June \citep{stappers14a}.  In both panels, the red line shows average $\gamma$-ray flux (with dotted red line showing the 1$\sigma$ errors) before and after the 2013 June state transition.
}
\label{fig:fermi}
\end{center}
\end{figure*}

\citet{stappers14a} showed that the $\gamma$-ray flux of J1023 increased by a factor of $\sim$5 in the LMXB state compared to the radio pulsar state, but were able to include only a few months of LMXB-state data. Our analysis of the longer span of data now available confirms their result, showing a sustained increase in the 0.1 -- 300 GeV flux by a factor of $6.5\pm 0.7$. Figure~\ref{fig:fermi} shows the $\gamma$-ray photon flux of J1023 in both the MSP and LMXB state; no trend or variability within the LMXB state is obvious.

\subsection{Comparison of the X-ray and radio light curves}
\begin{figure*}
\begin{center}
\includegraphics[width=0.85\textwidth]{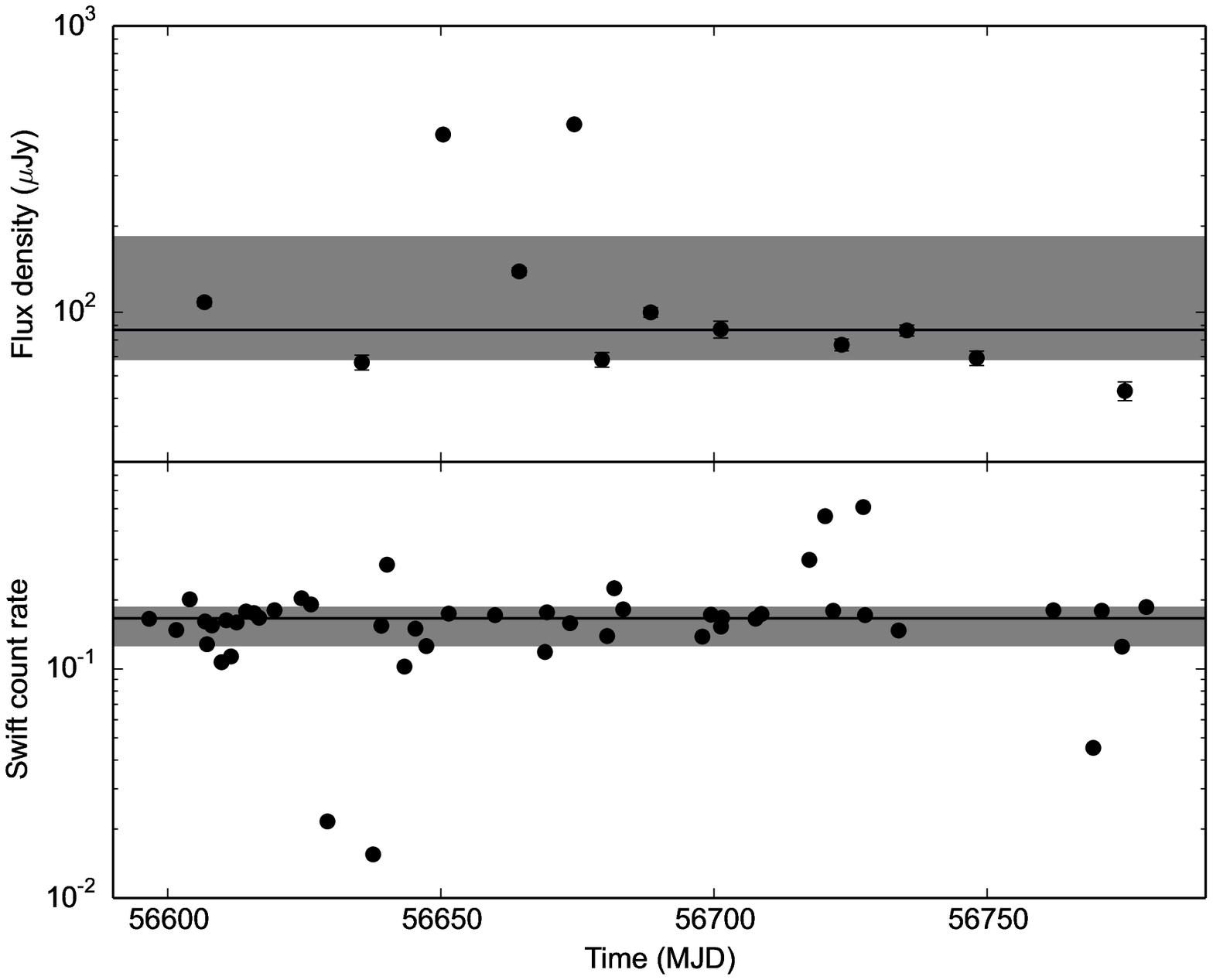}  
\caption{The X-ray and radio light curves from the VLA and \emph{Swift}, where the median value of each observation is plotted as a single point.  \emph{Swift} observations that were shorter than 15 minutes and hence unlikely to sample a representative range of the X-ray emission were excluded from the plot.  For both the radio and X-ray panels, the central 67\% range is shown as the gray region.  A few outlier points notwithstanding, the median X-ray luminosity remains exceedingly stable, within approximately $\pm20$\%.  
The radio is somewhat more variable, which may be due to the fact that the individual radio measurements generally required longer averages (up to 480 seconds, compared to 50 seconds for the {\em Swift} samples).  This yields a smaller number of time samples in the radio lightcurve, a considerable fraction of which would include times corresponding to two different X-ray modes averaged together, whereas the X-ray observations have more time points (with a cleaner separation between modes).  Since a 45-minute radio observation would sample a number of mode changes, this could lead to an increased scatter for the median radio points.}
\label{fig:mediancomparison}
\end{center}
\end{figure*}

In Figure~\ref{fig:mediancomparison}, we show the median luminosity of J1023 at each epoch in the radio and X-ray bands.  The median flux density for both the X-ray and radio emission remains remarkably constant over time, indicating that the system is in a relatively stable configuration during the LMXB state (on timescales  of months to over a year), as noted by \citet{archibald15a} and \citet{bogdanov15b}.  

The lack of sufficient simultaneity between the radio and X-ray observations prevents us from drawing conclusions as to whether the variability is correlated.  There are no simultaneous VLA and \emph{Swift} data, and only a 15 minute overlap between the first VLA observation and the aforementioned \emph{XMM-Newton} observation \citep[discussed separately by][]{bogdanov15b}.  For the purposes of the discussion below, we do not assume any correlation between the radio and X-ray variations and simply use the median value of radio luminosity and the median value of X-ray luminosity, along with error bars encompassing the most compact 67\% interval for both quantities.

\section{Discussion}
\label{sec:discussion}

\subsection{A jet origin for the radio emission of J1023}
\label{sec:disc:jet}
The radio flux density and spectral index, and the time variability thereof, seen for J1023 are indicative of synchrotron emission originating in material outflowing from the system.   No evidence is seen from the radio spectral index (down to timescales of 5 minutes) of steep spectrum emission that would be indicative of radio pulsar emission.  We cannot categorically rule out intermittent bursts of radio pulsar activity, but such periods would have to be both sparse and brief in order to remain consistent with our results, or else completely absorbed by free-free absorption by dense ionized material surrounding the neutron star.  Taken together with X-ray pulsations reported by \citet{archibald15a} that indicate near-continual accretion onto the neutron star surface, and the lack of radio pulsation detections reported in \citet{bogdanov15b}, the ubiquitous flat-spectrum radio emission makes an (even intermittently) active radio pulsar in J1023 in the present LMXB state appear unlikely.

In analogy to other LMXB systems, the most obvious interpretation of our data is that J1023 in its present state hosts a compact, partially self-absorbed jet powered by the accretion process.  The presence of collimated outflows has been previously been directly confirmed via high-resolution imaging in a number of cases for both neutron star and black hole LMXBs \citep{dhawan00a,stirling01a,fomalont01a,fender04a}.   
In the case of black hole systems, the evidence for jets is compelling even down to extremely low accretion rates: V404 Cygni is believed to host a jet in quiescence with $L_X \sim 10^{32.5}$ erg s$^{-1}$ \citep[e.g.,][]{hynes09a,miller-jones08a}, while A0620-00  \citep{gallo06a,russell13a} and XTE J1118+480 \citep{gallo14a,russell13a} also show radio emission at $L_X < 10^{32}$ erg s$^{-1}$ and evidence supportive of a jet break between the radio and optical bands \citep{gallo07a}.
For neutron star LMXBs, however, direct evidence of jets \citep[and other supporting evidence such as optical/near-infrared excesses that are most consistent with jet emission;][]{russell07a} has come only from systems that are undergoing accretion at much higher rates.  The X-ray luminosity of J1023 is more than two orders of magnitude lower than for any other neutron star system for which a jet had previously been inferred \citep{migliari11a}.

Alternative explanations for the flat spectral index ($-0.3 \lesssim \alpha \lesssim 0.3$) and high brightness temperature limit ($\mathrm{T}_{\mathrm{b}} \simeq 3 \times 10^{8}$\,K) for J1023 are, however, difficult to imagine.  Optically thin free-free emission could provide $\alpha\sim-0.1$, but would require an extremely hot population of electrons to match the observed brightness temperature, as $T_e = T_b/\tau$ and in the optically thin regime $\tau \ll 1$.  For optically thin synchrotron emission, obtaining a flat spectral index $\alpha\sim0$ requires an extremely hard electron population, with a power-law distribution index $p \sim 1$.  Such a hard distribution of electrons would be strongly at odds with the commonly assumed process of shock acceleration, which produces $2 < p < 2.5$ \citep{jones91a}.  We consider both of these scenarios (optically thin free-free emission and optically thin synchrotron emission) to be unlikely, and note that definitive proof of a jet in the J1023 system could be provided in the future by resolved VLBI imaging with a very sensitive global VLBI array.  Other strong, albeit not definitive, supporting evidence for the jet interpretation could come from the detection of a frequency dependent time delay in the J1023 lightcurve (where the magnitude of the delay implied a relativistic velocity), the detection of linear polarization (implying synchrotron emission), or the detection of a spectral break.

\subsection{A possible radio/X-ray correlation for transitional MSP systems}

\begin{figure*}
\begin{center}
\includegraphics[width=0.85\textwidth]{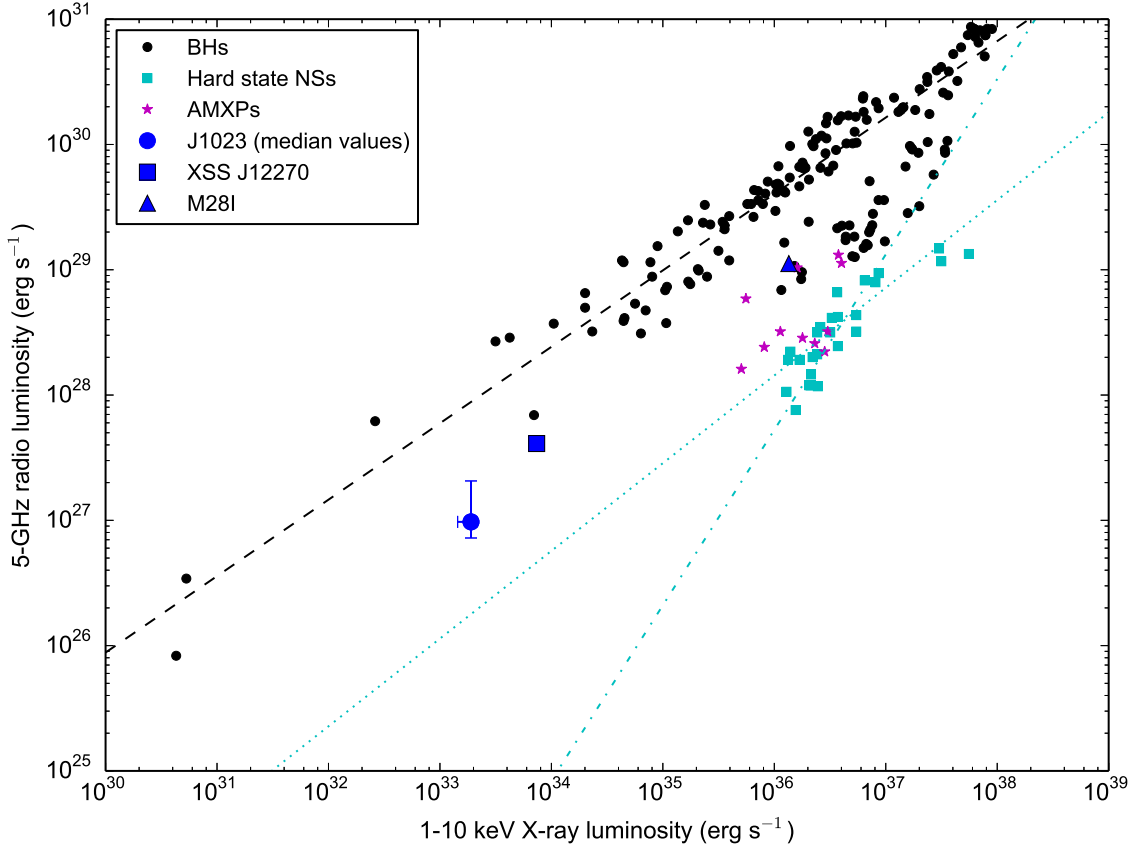}  
\caption{Radio and X-ray luminosity of known LMXB systems, including black hole systems (small black points) and neutron star systems (small magenta stars for AMXPs, small cyan squares for hard-state LXMBs, and large blue symbols for transitional MSP systems).  The best-fit correlation for the black hole systems from \citet{gallo14a} is shown with a black dashed line, and the ``atoll/Z-source" and ``hard-state" correlations for the neutron star systems presented in \citet{migliari06a} and \citet{migliari11a} are shown in cyan dotted and cyan dash-dot respectively.  J1023 (blue circle symbol) and the other two known transitional MSP systems XSS J12270 \citep[blue square symbol;][]{de-martino13a,bassa14a} and M28I \citep[blue triangle symbol;][]{papitto13a} are highlighted.  For J1023, we show an error box that spans the 67\% range of luminosity seen in our radio and X-ray monitoring (taking the median value from each observation, as shown in Figure~\ref{fig:mediancomparison}); the limited radio monitoring for XSS J12270 and M28I precludes such error bars for these sources, but we note that for M28I at least variability was seen within the single radio observation \citep{papitto13a}, and so similar uncertainties could be expected for these sources.  The transitional MSP systems appear to exhibit a similar scaling of radio luminosity with X-ray luminosity to the black hole systems, but sit in between the black hole systems and the neutron star systems.}
\label{fig:lrlx}
\end{center}
\end{figure*}

In Figure~\ref{fig:lrlx}, we show the radio and X-ray luminosity of J1023 along with the two other known transitional MSP systems XSS J12270 \citep{de-martino13a,bassa14a} and M28I \citep{papitto13a} and most of the published black hole and neutron star LMXB systems \citep{gallo06a,corbel08a,brocksopp10a,soleri10a,coriat11a,migliari11a,jonker12a,ratti12a,russell12a,corbel13a,gallo14a}.  
The radio luminosity shown in Figure~\ref{fig:lrlx} is defined as $L_R = 4 \pi d^2 \nu S_\nu$, where $d$ is the distance, $\nu$ is the reference radio frequency (in this case, 5 GHz) and $S_\nu$ is the flux density at this frequency.  Where necessary, a flat spectral index is assumed to convert measured flux densities to the reference frequency of 5 GHz; for J1023, we take the median 10 GHz flux density of 87 $\mu$Jy and use a flat spectral index (as noted in Section~\ref{sec:specindex}, the median spectral index is very close to 0).

The dashed line shows the well-established correlation between radio and X-ray luminosity for black holes \citep[$L_R \propto L_X^{0.61}$,][black line]{gallo14a}, as well as the two postulated correlations for neutron star systems ($L_R \propto L_X^{0.7}$, ``atoll/Z-source" correlation [blue dotted line]; \citealp{migliari06a}, and $L_R \propto L_X^{1.4}$, ``hard-state" correlation [blue dash-dotted line]; \citealp{migliari11a}).  We note that studies have differed on the exact correlation index for black hole LMXBs, finding values ranging between 0.6 and 0.7 \citep{gallo03a,gallo12a,gallo14a} and that a tighter correlation between radio and X-ray luminosity is generally seen for individual sources compared to the population as a whole \citep{gallo14a}.  Further, some black hole systems on the so-called ``outlier track'' (visible in Figure~\ref{fig:lrlx} as the cluster of X-ray underluminous points at $10^{36} \lesssim L_X \lesssim 10^{37.5}$\, erg s$^{-1}$, which extend down towards the neutron star systems) have been seen to exhibit a correlation index of 1.4 in some cases, which could be taken as evidence that some black hole LMXBs enter a radiatively efficient accretion regime under certain circumstances \citep{coriat11a}.  The discussion that follows focuses on lower accretion rates, where no evidence for radiatively efficient accretion in black hole systems has been seen, and is insensitive to changes in the correlation index for black hole LMXBs in the range 0.6 -- 0.7.

The three transitional MSP systems appear to be overluminous in the radio band (or underluminous in the X-ray band) compared to either postulated correlation for the neutron star systems; they appear to follow the same relationship as the black hole sources and the ``atoll/Z" neutron star sources, but with a radio luminosity approximately a factor of 5 fainter than the black hole sources, and an order of magnitude brighter than the ``atoll/Z" neutron star sources.  M28I occupies a point on the $L_R / L_X$ space that is encompassed by the AMXP population (the magenta stars in Figure~\ref{fig:lrlx}); AMXPs appear to be on average radio-overluminous when compared to other neutron star LMXBs of similar X-ray luminosity \citep{migliari11a}. Thus, the transitional MSP sources are clearly distinct from the hard-state neutron star LMXBs, and the AMXPs appear to straddle the two classes.  Below, we examine what processes could lead to this observational differentiation.

\subsection{Jet-dominated states in LMXB systems}

A variety of jet production mechanisms could potentially operate in LMXB systems.  \citet{blandford82a} and \citet{blandford77a} describe two models which have been widely applied to black hole systems, in which the disk magnetic field enables the collimation of the material emitted in the jet.  The former model, in which the jet is collimated by magnetic field lines anchored in the accretion disk, can be applied directly to neutron star systems.  The latter model extracts energy from the black hole spin and requires that disk magnetic field lines be anchored to the black hole ergosphere; as such it cannot be directly applied to neutron star systems, but the coupling of material in an accretion disk to a neutron star magnetosphere can still drive collimated outflows via a ``propeller" mode \citep[e.g.,][]{romanova05a,
ustyugova06a,romanova09a}.

Where radio emission in an LMXB system is thought to originate in a jet, it can be used as a proxy for the jet power.  For typical jet models, the observed radio luminosity $L_R$ is related to the total jet power $L_j$ by $L_R \propto L_j^{1.4}$ \citep[e.g.][]{blandford79a,markoff01a,falcke96a}.  The observed scaling of $L_R \propto L_X^{0.7}$ highlighted in Figure~\ref{fig:lrlx} for black hole LMXBs and (potentially) for transitional MSPs then implies that $L_j \propto L_X^{0.5}$.  If the accretion power $L_{\mathrm{tot}}$ is assumed to be liberated only via radiation (traced by $L_X$) or the jet ($L_j$), then we have $L_{\mathrm{tot}} = L_X + L_j$, and, by substitution, $L_{\mathrm{tot}} = L_X + A L_X^{0.5}$ \citep{fender03a}.  A boundary between two regimes occurs when $L_X = L_j = A L_X^{0.5}$; at higher accretion rates, the accretion is radiatively efficient and dominated by the X-ray output ($L_X \sim L_{tot}$, $L_X \propto \dot{M}$, and $L_j \propto \dot{M}^{0.5}$), while below this value, the accretion is radiatively inefficient and the system is jet-dominated ($L_X \ll L_{tot}$, $L_X \propto \dot{M}^2$, and $L_j \propto \dot{M}$).  For a number of black hole LMXBs, the coefficient $A$ has been estimated to range between 0.006 and 0.1 (where all luminosities are given in Eddington units), based on quasi-simultaneous measurements of $L_j$ and $L_X$ at a given accretion rate \citep{fender01a,corbel02a,fender03a,migliari06a}.  A key reason for this wide range of allowed values for $A$ is uncertainty in the ratio of radiated luminosity to total luminosity in the jet, as only the radiated component can be directly measured -- this ratio is generally assumed to be $\lesssim$0.05 \citep[e.g.][]{fender01a}, but remains poorly constrained.  Proceeding under the assumption that these values of $A$ are representative for all black hole LMXB systems leads to the determination that all quiescent black hole LMXBs are accreting below the threshold level and hence in a jet-dominated state \citep{fender03a,migliari06a}, although with a sufficiently low value of $A$ it would be possible for a system to remain radiatively efficient even down to quiescent luminosities.  The potential for advection of energy across the black hole event horizon complicates this simple interpretation, adding a new ``sink" for some fraction of $L_{\mathrm{tot}}$ and meaning that a system could be radiatively inefficient and yet not jet-dominated; however, this process cannot operate in the neutron star LMXBs.

If the difference in radio luminosity between neutron star LMXBs and black hole LMXBs is due purely to a difference in jet power, then the representative values of $A$ calculated for black hole LMXBs can be scaled to calculate the accretion rate at which neutron star LMXBs will transition into a radiatively inefficient, jet-dominated regime.  Measurements of neutron star LMXBs at higher accretion rates (small cyan squares in Figure~\ref{fig:lrlx}) show an average radio luminosity 30 times below that given by the black hole $L_R / L_X$ relationship, meaning this transition would occur around or below the lowest accretion rates inferred based on X-ray measurements \citep{fender03a}.  Such an interpretation {\em assumes} a scaling of $L_R \propto L_X^{0.7}$ that holds down to quiescence, since previous observations have not been sensitive enough to make radio detections at lower accretion rates.  For systems which exhibit $L_R \propto L_X^{1.4}$, as has been suggested by \citet{migliari03a} for 4U 1728--34 and shown by the cyan dash-dot line in Figure~\ref{fig:lrlx}, then the system will remain radiatively efficient at all accretion rates ($L_j \propto L_X \propto \dot{m}$; $L_j < L_X$).

The higher-than-expected radio luminosity of the 3 transitional MSP systems indicates that they become radiatively inefficient and jet-dominated at low (but observable) accretion rates.  The factor-of-5 difference in radio luminosity compared to the black hole LMXBs implies a coefficient $A$ for the transitional MSP systems that is only 3 times lower than that of the black hole LMXBs, meaning that the transition to radiatively inefficient accretion occurs at an accretion rate 10 times lower (in Eddington units).  Assuming that for the black hole systems $A$ lies somewhere in the range 0.006 to 0.1 as described above, this corresponds to an X-ray luminosity of $7\times10^{33}$ - $2\times10^{36}$ erg s$^{-1}$ for a 1.4 M$_{\odot}$ neutron star.  J1023, with a 1--10 keV luminosity of $2\times10^{33}$ erg s$^{-1}$, appears therefore to be  jet-dominated.  XSS J12270 is likewise very probably jet-dominated, but in contrast, the observations of M28I at $\gtrsim1\times10^{36}$ erg s$^{-1}$ (around 1\% of the Eddington luminosity) are more likely to be in the radiation-dominated, radiatively efficient regime.

There are a number of influences on the radiative efficiency of the accretion flow and radiative efficiency and power of the jet that could lead to the observed offset between the transitional MSPs and the black hole LMXBs in the $L_R - L_X$ correlation.
 First, the mass of the compact object is likely to affect the jet properties.  The ``fundamental plane" relationship for accreting black holes shows, for a fixed X-ray luminosity, a dependence of radio luminosity on the black hole mass $M_{\mathrm{BH}}$ of the form $L_R \propto M_{\mathrm{BH}}^{\beta}$, where $\beta \simeq 0.6-0.8$  \citep{merloni03a,falcke04a,plotkin12a}.  Based on this relationship, a ``typical" stellar-mass black hole of $\sim$8 M$_{\odot}$ \citep{kreidberg12a} would already be 3--4 times more radio-luminous than a neutron-star mass (1.4 M$_{\odot}$) black hole with the same X-ray luminosity.  Interaction with the neutron star magnetic field could also affect the efficiency of jet formation.  The radius at which the accretion disk transitions to a radiatively inefficient flow (at a given accretion rate) could differ, altering the radiative efficiency of the disk.  Finally, the small amount of material that reaches the neutron star surface will also contribute to the radiated energy in the transitional MSP case, whereas this energy could be removed by advection in the case of a black hole LMXB.  The exact contribution of material accreting on to the neutron star surface to the X-ray luminosity of J1023 is unclear; only in the high mode do we have a lower limit of 8\% from the detection of coherent pulsations \citep{archibald15a,bogdanov15b}.   This value is a lower limit because the total X-ray emission originating from the hotspots will exceed the observed 8\% pulsed fraction, due to the finite hotspot size and propagation effects such as gravitational bending \citep{beloborodov02a}.

\subsection{The influence of propellor-mode accretion}

We speculate that the (at least intermittent) activation of a
propeller-style accretion mode is a key reason for the apparently
higher-efficiency jet formation in the transitional MSP systems, compared to other neutron star LMXBs.  This is an attractive proposition in
that 1) it naturally explains the transition to a jet-dominated regime
at lower accretion rates (and hence X-ray luminosities), since the
propeller-mode can only operate at these lower accretion rates
\citep{ustyugova06a}, 2) simulations have shown that the
high-velocity, low-density, collimated axial outflow dominates the
energetics of the expelled material \citep{romanova09a}, and 3) a
large majority, but not all, of the material is ejected, leaving a
small fraction able to reach the neutron star surface
\citep{romanova09a,ustyugova06a}.  This offers the possibility of explaining the
X-ray pulsations observed by \citet{archibald15a} and the
radio emission reported here, as well as the similar results for XSS J12270 \citep{papitto15a}.  However, we caution that the 
results of \citet{romanova09a} are based exclusively on 
magnetohydrodynamic modeling, and so the radiative processes
that could lead to the observed radio emission still need to be carefully 
modeled.

Finally, propeller-mode accretion could also explain the high-energy properties of J1023.  The marked increase in $\gamma$-ray luminosity in the LMXB state seen by \citet{stappers14a}, and confirmed here to be roughly constant throughout the LMXB state, has previously been hypothesized to be powered by a pulsar wind interaction \citep{stappers14a,coti-zelati14a}.  Based on our radio imaging data, however, we judge it unlikely that the radio pulsar ever becomes active during the current LMXB state.  In the propeller-mode accretion scenario, the $\gamma$-ray luminosity increase can instead be explained by synchrotron self-Compton emission from electrons accelerated at the inner edge of the accretion disk by the pulsar's magnetic field.  Such a model was developed by \citet{papitto14a} to explain the high-energy emission of XSS J12270; if it is correct, and if (as we predict) all transitional MSP objects undergo propeller-mode accretion at low accretion rates, then it is to be expected that J1023 would produce $\gamma$-ray emission via the same mechanism.  

However, also in this case there are a few caveats to consider. The model proposed by \citet{papitto14a} assumes the presence of a Fermi acceleration process occurring at the disk/magnetosphere boundary, which, although plausible, has not been previously demonstrated to occur in LMXB systems. Furthermore, as noted by \citet{papitto14a}, the emission region responsible for the high-energy emission cannot also produce the observed radio emission; as we have shown, the size of the radio-emitting region in J1023 must be of order 0.5--30 times the binary separation of 4.3 light-seconds.
Finally, the \citet{papitto14a} model does not include accretion onto the neutron star surface, although they note that this is a possibility if the amount of inflowing material is very small.  The X-ray pulsations seen by \citet{archibald15a} for J1023 show that at least some of the X-ray luminosity of J1023 \emph{is} powered by accretion onto the neutron star surface   After submission of this manuscript, the discovery of X-ray pulsations with similar properties for XSS J12270 was also published \citep{papitto15a}, reinforcing the necessity of considering the flow of material to the neutron star surface.  Accordingly, the model of \citet{papitto14a} would need further development to reach a self-consistent description of the properties of J1023 from optical to $\gamma$-rays, and further extension (ideally incorporating radiative modeling based on the magnetohydrodynamic models of \citealp{romanova09a}) to cover the radio emission.  Given the wealth of multiwavelength data that has been assembled for J1023 \citep{bogdanov15b}, such a detailed analysis should be feasible in a future work. 

An analysis of the accretion physics of J1023 is complicated by the fact that J1023 has been seen to switch between three distinct X-ray luminosity modes while in its current LMXB state \citep[the ``low", ``high", and ``flare" modes;][]{archibald15a,bogdanov15b} on timescales ranging from 10s of seconds to hours.  Very similar X-ray modality has also recently been seen in XSS J12270 \citep{papitto15a}.  The jet production mechanism in J1023 might operate in only one of the three X-ray modes, or in several or indeed all modes; alternatively, a jet may be present in all modes but with substantially changing properties.  As with the radio emission, the $\gamma$-ray production could differ between the observed modes.  Useful extra information could be provided by truly contemporaneous and high-sensitivity radio and X-ray observations, allowing a cross-correlation analysis of the radio and X-ray light curves.  If a time delay and smoothing time for the radio emission could be measured, this could provide useful information on the jet velocity, the size of the radio emitting region, and its separation from the neutron star.  Resolving the jet and tracking the motion of jet components using VLBI imaging (ideally in conjunction with X-ray observations) would offer a different, similarly powerful handle on the jet physics; however, as our EVN observations showed, this would require an extremely sensitive VLBI array, as well the good fortune to observe during a period of above-average radio luminosity.


\subsection{Expanding the study of jets in neutron star LMXBs at low accretion rates}

In addition to ongoing investigation of J1023 in the LMXB state, the identification of additional members of this apparent source class of jet-dominated neutron star LMXBs would be extremely helpful in furthering our understanding of the accretion process(es) in this regime.  This is especially true since the current single $L_R/L_X$ data points for J1023 and XSS J12270 cannot reveal the correlation index for the individual sources, whereas several black hole LMXBs have had measurements made over several orders of magnitude in X-ray luminosity (and hence accretion rate) which allow this estimation to be made for individual sources in addition to the population as a whole \citep{gallo14a}.
Several other known LMXBs display X-ray properties more or less similar to the transitional MSP systems: variable X-ray luminosities, which cluster above their quiescent values at around $10^{33}$ erg s$^{-1}$ for extended periods of time, and a spectrum with a hard power-law component ($\Gamma < 2$).  A summary of such known systems is given in \citet{degenaar14a}.  However, with one exception, all are considerably more distant than J1023, making the task of identifying any flat-spectrum radio emission challenging.


The nearest and most promising source is Cen X$-$4 \citep[$d \sim 1.2$\,kpc;][]{chevalier89a}.  Whilst the X-ray luminosity and variability are similar to J1023, \citet{chakrabarty14a} and \citet{dangelo15a} show that the X-ray spectrum is considerably different, with a significant thermal component and a cutoff of the power-law component above 10 keV.  In contrast, \citet{tendulkar14a} find no such cutoff for J1023.  \citet{bernardini13a} suggest that the X-ray emission of Cen X$-$4 is derived primarily from accretion onto the neutron-star surface and find it unlikely that a strong, collimated outflow would be present.  Given the proximity of Cen X-4, confirming or denying the presence of an outflow should be relatively straightforward with a combined radio/X-ray observing campaign.  Other sources include XMM J174457-2850.3 \citep[$d = 6.5$\,kpc;][]{degenaar14a}, SAX J1808.4-3658 \citep[a known AMXP at $d=2.5-3.5$\,kpc that has been detected several times in the radio during outburst;][]{gaensler99a,rupen05a,galloway06a,campana08a}, EXO 1745$-$248 \citep[$d\sim8.7$\,kpc][]{wijnands05a}, and Aql X$-$1 \citep[an atoll source at $d\sim5$\,kpc which has once shown coherent X-ray pulsations and has been detected in radio continuum at higher luminosities;][]{rutledge02a,casella08a,tudose09a}.  However, whilst all of these are strong candidates for propeller-mode accretion, their distance means that the expected average radio flux density would be of the order 3 -- 25 $\mu$Jy, making a detection challenging to near-impossible, even with very deep VLA imaging.  The identification of new transitional MSP systems (either found in the LMXB state, such as the system described by \citealt{bogdanov15a}, or objects from the known population of ``redback" or ``black widow" pulsars which transition to the LMXB state) is an alternate possibility for providing new data points on the radio/X-ray correlation.  A third option is a radio survey of other neutron star LMXB systems which display a hard X-ray spectrum at intermediate X-ray luminosities ($10^{34} < L_X < 10^{35}$ erg s$^{-1}$), as the radio emission should also be brighter and hence easier to detect to large distances in a reasonable observing time.  Some candidate intermediate luminosity sources are listed in \citet{degenaar14a}.

\subsection{Flat-spectrum radio sources in globular clusters}

Finally, we consider the implications of this result for observations of accreting sources in globular clusters.  As shown in Figure~\ref{fig:lrlx}, the three known transitional MSP systems all exhibit higher radio luminosity than models based on neutron star binaries with higher accretion rates had predicted.  Rather than being two or more orders of magnitude fainter in the radio band than black hole systems with an equivalent X-ray luminosity, the difference appears to be closer to a factor of five, which is comparable to the scatter around the correlation (in the case of the black hole correlation; the statistics for the transitional MSP sources are too low to estimate the scatter).  This relative radio brightness invites a re-examination of globular cluster black hole candidates in M22 \citep{strader12a} and M62 \citep{chomiuk13a}, where the ratio of X-ray to radio luminosity was a key element in the interpretation.

The M62 candidate source falls exactly on the $L_R - L_X$ correlation for black holes, and so a black hole interpretation must still certainly be favoured.  However, if the scatter around the transitional MSP correlation is comparable to that of the black hole correlation, then from the point of view of the X-ray and radio luminosity it is certainly plausible that the M62 source is actually a transitional MSP.  The likely association of a giant star as the companion object \citep{chomiuk13a} would, however, mean that the orbital period of the system would be longer than that of a typical ``redback" MSP.  Of order 10 redback MSPs are known in globular clusters\footnote{A list of known redback pulsars is maintained at \url{http://www.naic.edu/\~{}pfreire/GCpsr.html}}, indicating the potential for transitional systems to be found.  In light of the rapid variability demonstrated by J1023, the fact that the radio and X-ray observations of M62 were not contemporaneous is particularly important, as it suggests that a single measurement of radio luminosity and a single measurement of X-ray luminosity may not be robust.

The two M22 sources fall well above the $L_R - L_X$ correlation for black holes (that is, they are even more radio bright than predicted by the best-fit correlation for black hole systems).  At face value, that argues against a transitional MSP interpretation.  However, as with the M62 candidate, the observations were not simultaneous, which offers both the potential for ``normal" variability or a change in the state of the system.  In this scenario, the radio observations of the source would have been carried out during the LMXB state, and the X-ray observations during the radio pulsar state.  While the probability of both sources changing state between the X-ray and radio observations is unlikely (with the caveat that our knowledge of how long transitional MSPs spend in the LMXB state per transition is limited), the fact that the sources fall quite far from the $L_R - L_X$ correlation for black hole LMXBs also requires explanation.  For both M22 and M62, truly simultaneous radio and X-ray observations (or a quasi-simultaneous, extended monitoring campaign such as undertaken for J1023) would give much more certainty about whether a transitional MSP system could be responsible.

\section{Conclusions}
We have monitored the transitional MSP J1023 in its current LMXB state over a period of 6 months, detecting variable X-ray emission and variable, flat-spectrum radio emission that is suggestive of a compact jet.  In both the radio and X-ray bands, the average luminosity over months to a year is quite stable, despite variability of two orders of magnitude in luminosity on timescales of seconds to hours.  When J1023 is compared to the other two known transitional MSPs and other black hole and neutron star LMXBs, it appears that the transitional MSPs exhibit a different correlation between radio and X-ray luminosity than either the black hole LMXB systems or the relationships previously proposed for neutron star LMXB systems accreting at higher rates \citep{migliari06a,migliari11a}.  The apparent correlation seen for the transitional MSP systems does, however, extend to a subset of the AMXP sources, which were already known to be the most radio-loud (for a given X-ray luminosity) of the neutron star LMXB systems \citep{migliari11a}.  We hypothesize that the transitional MSPs are jet-dominated accretion systems operating in a propeller mode, where only a fraction of the mass transferred from the secondary reaches the neutron star surface and the majority is ejected in a jet.  The similarity to AMXP systems implies that they, too, may become jet-dominated in some cases.  In this scenario, the $\gamma$-ray emission seen from J1023 in the LMXB state originates from the acceleration of particles at the inner edge of the accretion disk by the pulsar magnetic field, and the radio emission is generated in the collimated outflow.  We predict that neutron star systems accreting in propeller-mode will generically show radio jets and $\gamma$-ray emission, and will follow a radio -- X-ray correlation of $L_R \propto L_X^{0.7}$, which will break down at a sufficiently high mass accretion rate (and hence X-ray luminosity), when accretion can no longer proceed via a mechanism where the majority of the material is expelled.  
It remains unclear whether the transition to a jet-dominated regime is a common occurrence for neutron star LMXBs at sufficiently low mass transfer rates, or if a property intrinsic to the transitional MSPs / AMXPs such as magnetic field strength and/or spin period is important in enabling jet formation.  The \citet{papitto14a} model of propeller-mode accretion which predicts $\gamma$-ray emission is strongly dependent on the neutron star period and magnetic field strength, and it is likely that these parameters are also important for the generation of the jet which powers the radio emission.  Deep radio observations of a wider range of neutron star LMXB systems in quiescence would be desirable to answer these questions.
Finally, we note that future searches for black holes in globular clusters should make use of contemporaneous radio and X-ray observations to distinguish black hole systems from transitional MSP systems similar to J1023.

\acknowledgements  We thank the referee, Rob Fender, for constructive suggestions which improved this manuscript.  ATD is supported by an NWO Veni Fellowship. JCAMJ is supported by an Australian Research Council (ARC) Future Fellowship (FT140101082) and also acknowledges support from an ARC Discovery Grant (DP120102393).  A.P. acknowledge support from an NWO Vidi fellowship. J.W.T.H. acknowledges funding from an NWO Vidi fellowship and ERC Starting Grant ``DRAGNET" (337062).  A.M.A. was funded for this work via an NWO Vrije Competitie grant (PI Hessels).   The National Radio Astronomy Observatory is a facility of the National Science Foundation operated under cooperative agreement by Associated Universities, Inc.   The EVN (\url{http://www.evlbi.org}) is a joint facility of European, Chinese, South African,
and other radio astronomy institutes funded by their national research councils. The WSRT is
operated by ASTRON (Netherlands Institute for Radio Astronomy) with support from the
Netherlands Foundation for Scientific Research. 
LOFAR, the Low Frequency Array designed and constructed by ASTRON,
  has facilities in several countries, that are owned by various
  parties (each with their own funding sources), and that are
  collectively operated by the International LOFAR Telescope (ILT)
  foundation under a joint scientific policy.
The research leading to these results has
received funding from the European Commission Seventh Framework Programme (FP/2007-2013)
under grant agreement No. 283393 (RadioNet3). e-VLBI research infrastructure in Europe was
supported by the European Union's Seventh Framework Programme (FP7/2007-2013) under grant
agreement number RI-261525 NEXPReS.  
This research has made use of NASA's Astrophysics Data System. 
 This research has made use of the NASA/IPAC Extragalactic Database (NED) which is operated by the Jet Propulsion Laboratory, California Institute of Technology, under contract with the National Aeronautics and Space Administration.
Funding for SDSS-III has been provided by the Alfred P. Sloan Foundation, the Participating Institutions, the National Science Foundation, and the U.S. Department of Energy Office of Science. The SDSS-III web site is http://www.sdss3.org/.

\bibliographystyle{apj}
\bibliography{deller_thesis}

\begin{thebibliography}{}
\expandafter\ifx\csname natexlab\endcsname\relax\def\natexlab#1{#1}\fi

\bibitem[{{Ahn} {et~al.}(2014){Ahn}, {Alexandroff}, {Allende Prieto}, {Anders},
  {Anderson}, {Anderton}, {Andrews}, {Aubourg}, {Bailey}, {Bastien}, \&
  et~al.}]{ahn14a}
{Ahn}, C.~P., {Alexandroff}, R., {Allende Prieto}, C., {et~al.} 2014, \apjs,
  211, 17

\bibitem[{{Alpar} {et~al.}(1982){Alpar}, {Cheng}, {Ruderman}, \&
  {Shaham}}]{alpar82a}
{Alpar}, M.~A., {Cheng}, A.~F., {Ruderman}, M.~A., \& {Shaham}, J. 1982, \nat,
  300, 728

\bibitem[{{Archibald} {et~al.}(2010){Archibald}, {Kaspi}, {Bogdanov},
  {Hessels}, {Stairs}, {Ransom}, \& {McLaughlin}}]{archibald10a}
{Archibald}, A.~M., {Kaspi}, V.~M., {Bogdanov}, S., {et~al.} 2010, \apj, 722,
  88

\bibitem[{{Archibald} {et~al.}(2013){Archibald}, {Kaspi}, {Hessels},
  {Stappers}, {Janssen}, \& {Lyne}}]{archibald13a}
{Archibald}, A.~M., {Kaspi}, V.~M., {Hessels}, J.~W.~T., {et~al.} 2013, ArXiv
  e-prints, arXiv:1311.5161

\bibitem[{{Archibald} {et~al.}(2009){Archibald}, {Stairs}, {Ransom}, {Kaspi},
  {Kondratiev}, {Lorimer}, {McLaughlin}, {Boyles}, {Hessels}, {Lynch}, {van
  Leeuwen}, {Roberts}, {Jenet}, {Champion}, {Rosen}, {Barlow}, {Dunlap}, \&
  {Remillard}}]{archibald09a}
{Archibald}, A.~M., {Stairs}, I.~H., {Ransom}, S.~M., {et~al.} 2009, Science,
  324, 1411

\bibitem[{{Archibald} {et~al.}(2015){Archibald}, {Bogdanov}, {Patruno},
  {Hessels}, {Deller}, {Bassa}, {Janssen}, {Kaspi}, {Lyne}, {Stappers},
  {Tendulkar}, {D'Angelo}, \& {Wijnands}}]{archibald15a}
{Archibald}, A.~M., {Bogdanov}, S., {Patruno}, A., {et~al.} 2015, \apj, 807, 62

\bibitem[{{Bassa} {et~al.}(2014){Bassa}, {Patruno}, {Hessels}, {Keane},
  {Monard}, {Mahony}, {Bogdanov}, {Corbel}, {Edwards}, {Archibald}, {Janssen},
  {Stappers}, \& {Tendulkar}}]{bassa14a}
{Bassa}, C.~G., {Patruno}, A., {Hessels}, J.~W.~T., {et~al.} 2014, \mnras, 441,
  1825

\bibitem[{{Beloborodov}(2002)}]{beloborodov02a}
{Beloborodov}, A.~M. 2002, \apjl, 566, L85

\bibitem[{{Bernardini} {et~al.}(2013){Bernardini}, {Cackett}, {Brown},
  {D'Angelo}, {Degenaar}, {Miller}, {Reynolds}, \& {Wijnands}}]{bernardini13a}
{Bernardini}, F., {Cackett}, E.~M., {Brown}, E.~F., {et~al.} 2013, \mnras, 436,
  2465

\bibitem[{{Blandford} \& {K{\"o}nigl}(1979)}]{blandford79a}
{Blandford}, R.~D., \& {K{\"o}nigl}, A. 1979, \apj, 232, 34

\bibitem[{{Blandford} \& {Payne}(1982)}]{blandford82a}
{Blandford}, R.~D., \& {Payne}, D.~G. 1982, \mnras, 199, 883

\bibitem[{{Blandford} \& {Znajek}(1977)}]{blandford77a}
{Blandford}, R.~D., \& {Znajek}, R.~L. 1977, \mnras, 179, 433

\bibitem[{{Bogdanov} \& {Halpern}(2015)}]{bogdanov15a}
{Bogdanov}, S., \& {Halpern}, J.~P. 2015, \apjl, 803, L27

\bibitem[{{Bogdanov} {et~al.}(2015){Bogdanov}, {Archibald}, {Bassa}, {Deller},
  {Halpern}, {Heald}, {Hessels}, {Janssen}, {Lyne}, {Mold{\'o}n}, {Paragi},
  {Patruno}, {Perera}, {Stappers}, {Tendulkar}, {D'Angelo}, \&
  {Wijnands}}]{bogdanov15b}
{Bogdanov}, S., {Archibald}, A.~M., {Bassa}, C., {et~al.} 2015, \apj, 806, 148

\bibitem[{{Bradley} {et~al.}(2007){Bradley}, {Hynes}, {Kong}, {Haswell},
  {Casares}, \& {Gallo}}]{bradley07a}
{Bradley}, C.~K., {Hynes}, R.~I., {Kong}, A.~K.~H., {et~al.} 2007, \apj, 667,
  427

\bibitem[{{Brocksopp} {et~al.}(2010){Brocksopp}, {Jonker}, {Maitra}, {Krimm},
  {Pooley}, {Ramsay}, \& {Zurita}}]{brocksopp10a}
{Brocksopp}, C., {Jonker}, P.~G., {Maitra}, D., {et~al.} 2010, \mnras, 404, 908

\bibitem[{{Campana} {et~al.}(2008){Campana}, {Stella}, \&
  {Kennea}}]{campana08a}
{Campana}, S., {Stella}, L., \& {Kennea}, J.~A. 2008, \apjl, 684, L99

\bibitem[{{Cantrell} {et~al.}(2008){Cantrell}, {Bailyn}, {McClintock}, \&
  {Orosz}}]{cantrell08a}
{Cantrell}, A.~G., {Bailyn}, C.~D., {McClintock}, J.~E., \& {Orosz}, J.~A.
  2008, \apjl, 673, L159

\bibitem[{{Casella} {et~al.}(2008){Casella}, {Altamirano}, {Patruno},
  {Wijnands}, \& {van der Klis}}]{casella08a}
{Casella}, P., {Altamirano}, D., {Patruno}, A., {Wijnands}, R., \& {van der
  Klis}, M. 2008, \apjl, 674, L41

\bibitem[{{Chakrabarty} {et~al.}(2014){Chakrabarty}, {Tomsick}, {Grefenstette},
  {Psaltis}, {Bachetti}, {Barret}, {Boggs}, {Christensen}, {Craig},
  {F{\"u}rst}, {Hailey}, {Harrison}, {Kaspi}, {Miller}, {Nowak}, {Rana},
  {Stern}, {Wik}, {Wilms}, \& {Zhang}}]{chakrabarty14a}
{Chakrabarty}, D., {Tomsick}, J.~A., {Grefenstette}, B.~W., {et~al.} 2014,
  \apj, 797, 92

\bibitem[{{Chevalier} {et~al.}(1989){Chevalier}, {Ilovaisky}, {van Paradijs},
  {Pedersen}, \& {van der Klis}}]{chevalier89a}
{Chevalier}, C., {Ilovaisky}, S.~A., {van Paradijs}, J., {Pedersen}, H., \&
  {van der Klis}, M. 1989, \aap, 210, 114

\bibitem[{{Chomiuk} {et~al.}(2013){Chomiuk}, {Strader}, {Maccarone},
  {Miller-Jones}, {Heinke}, {Noyola}, {Seth}, \& {Ransom}}]{chomiuk13a}
{Chomiuk}, L., {Strader}, J., {Maccarone}, T.~J., {et~al.} 2013, \apj, 777, 69

\bibitem[{{Corbel} {et~al.}(2013){Corbel}, {Coriat}, {Brocksopp}, {Tzioumis},
  {Fender}, {Tomsick}, {Buxton}, \& {Bailyn}}]{corbel13a}
{Corbel}, S., {Coriat}, M., {Brocksopp}, C., {et~al.} 2013, \mnras, 428, 2500

\bibitem[{{Corbel} \& {Fender}(2002)}]{corbel02a}
{Corbel}, S., \& {Fender}, R.~P. 2002, \apjl, 573, L35

\bibitem[{{Corbel} {et~al.}(2000){Corbel}, {Fender}, {Tzioumis}, {Nowak},
  {McIntyre}, {Durouchoux}, \& {Sood}}]{corbel00a}
{Corbel}, S., {Fender}, R.~P., {Tzioumis}, A.~K., {et~al.} 2000, \aap, 359, 251

\bibitem[{{Corbel} {et~al.}(2008){Corbel}, {Koerding}, \& {Kaaret}}]{corbel08a}
{Corbel}, S., {Koerding}, E., \& {Kaaret}, P. 2008, \mnras, 389, 1697

\bibitem[{{Coriat} {et~al.}(2011){Coriat}, {Corbel}, {Prat}, {Miller-Jones},
  {Cseh}, {Tzioumis}, {Brocksopp}, {Rodriguez}, {Fender}, \&
  {Sivakoff}}]{coriat11a}
{Coriat}, M., {Corbel}, S., {Prat}, L., {et~al.} 2011, \mnras, 414, 677

\bibitem[{{Coti Zelati} {et~al.}(2014){Coti Zelati}, {Baglio}, {Campana},
  {D'Avanzo}, {Goldoni}, {Masetti}, {Mu{\~n}oz-Darias}, {Covino}, {Fender},
  {Jim{\'e}nez Bail{\'o}n}, {Ot{\'{\i}}-Floranes}, {Palazzi}, \&
  {Ram{\'o}n-Fox}}]{coti-zelati14a}
{Coti Zelati}, F., {Baglio}, M.~C., {Campana}, S., {et~al.} 2014, \mnras, 444,
  1783

\bibitem[{{D'Angelo} {et~al.}(2015){D'Angelo}, {Fridriksson}, {Messenger}, \&
  {Patruno}}]{dangelo15a}
{D'Angelo}, C.~R., {Fridriksson}, J.~K., {Messenger}, C., \& {Patruno}, A.
  2015, \mnras, 449, 2803

\bibitem[{{D'Angelo} \& {Spruit}(2010)}]{dangelo10a}
{D'Angelo}, C.~R., \& {Spruit}, H.~C. 2010, \mnras, 406, 1208

\bibitem[{{D'Angelo} \& {Spruit}(2012)}]{dangelo12a}
---. 2012, \mnras, 420, 416

\bibitem[{{de Martino} {et~al.}(2013){de Martino}, {Belloni}, {Falanga},
  {Papitto}, {Motta}, {Pellizzoni}, {Evangelista}, {Piano}, {Masetti},
  {Bonnet-Bidaud}, {Mouchet}, {Mukai}, \& {Possenti}}]{de-martino13a}
{de Martino}, D., {Belloni}, T., {Falanga}, M., {et~al.} 2013, \aap, 550, A89

\bibitem[{{Degenaar} {et~al.}(2014){Degenaar}, {Wijnands}, {Reynolds},
  {Miller}, {Altamirano}, {Kennea}, {Gehrels}, {Haggard}, \&
  {Ponti}}]{degenaar14a}
{Degenaar}, N., {Wijnands}, R., {Reynolds}, M.~T., {et~al.} 2014, \apj, 792,
  109

\bibitem[{{Deller} {et~al.}(2012){Deller}, {Archibald}, {Brisken},
  {Chatterjee}, {Janssen}, {Kaspi}, {Lorimer}, {Lyne}, {McLaughlin}, {Ransom},
  {Stairs}, \& {Stappers}}]{deller12b}
{Deller}, A.~T., {Archibald}, A.~M., {Brisken}, W.~F., {et~al.} 2012, \apjl,
  756, L25

\bibitem[{{Dhawan} {et~al.}(2000){Dhawan}, {Mirabel}, \&
  {Rodr{\'{\i}}guez}}]{dhawan00a}
{Dhawan}, V., {Mirabel}, I.~F., \& {Rodr{\'{\i}}guez}, L.~F. 2000, \apj, 543,
  373

\bibitem[{{Ek{\c s}i} \& {Alpar}(2005)}]{eksi05a}
{Ek{\c s}i}, K.~Y., \& {Alpar}, M.~A. 2005, \apj, 620, 390

\bibitem[{{Falcke} \& {Biermann}(1996)}]{falcke96a}
{Falcke}, H., \& {Biermann}, P.~L. 1996, \aap, 308, 321

\bibitem[{{Falcke} {et~al.}(2004){Falcke}, {K{\"o}rding}, \&
  {Markoff}}]{falcke04a}
{Falcke}, H., {K{\"o}rding}, E., \& {Markoff}, S. 2004, \aap, 414, 895

\bibitem[{{Fender} {et~al.}(1998){Fender}, {Spencer}, {Tzioumis}, {Wu}, {van
  der Klis}, {van Paradijs}, \& {Johnston}}]{fender98a}
{Fender}, R., {Spencer}, R., {Tzioumis}, T., {et~al.} 1998, \apjl, 506, L121

\bibitem[{{Fender} {et~al.}(2004){Fender}, {Wu}, {Johnston}, {Tzioumis},
  {Jonker}, {Spencer}, \& {van der Klis}}]{fender04a}
{Fender}, R., {Wu}, K., {Johnston}, H., {et~al.} 2004, \nat, 427, 222

\bibitem[{{Fender}(2001)}]{fender01a}
{Fender}, R.~P. 2001, \mnras, 322, 31

\bibitem[{{Fender} {et~al.}(2003){Fender}, {Gallo}, \& {Jonker}}]{fender03a}
{Fender}, R.~P., {Gallo}, E., \& {Jonker}, P.~G. 2003, \mnras, 343, L99

\bibitem[{{Fomalont} {et~al.}(2001){Fomalont}, {Geldzahler}, \&
  {Bradshaw}}]{fomalont01a}
{Fomalont}, E.~B., {Geldzahler}, B.~J., \& {Bradshaw}, C.~F. 2001, \apj, 558,
  283

\bibitem[{{Gaensler} {et~al.}(1999){Gaensler}, {Stappers}, \&
  {Getts}}]{gaensler99a}
{Gaensler}, B.~M., {Stappers}, B.~W., \& {Getts}, T.~J. 1999, \apjl, 522, L117

\bibitem[{{Gallo} {et~al.}(2006){Gallo}, {Fender}, {Miller-Jones}, {Merloni},
  {Jonker}, {Heinz}, {Maccarone}, \& {van der Klis}}]{gallo06a}
{Gallo}, E., {Fender}, R.~P., {Miller-Jones}, J.~C.~A., {et~al.} 2006, \mnras,
  370, 1351

\bibitem[{{Gallo} {et~al.}(2003){Gallo}, {Fender}, \& {Pooley}}]{gallo03a}
{Gallo}, E., {Fender}, R.~P., \& {Pooley}, G.~G. 2003, \mnras, 344, 60

\bibitem[{{Gallo} {et~al.}(2007){Gallo}, {Migliari}, {Markoff}, {Tomsick},
  {Bailyn}, {Berta}, {Fender}, \& {Miller-Jones}}]{gallo07a}
{Gallo}, E., {Migliari}, S., {Markoff}, S., {et~al.} 2007, \apj, 670, 600

\bibitem[{{Gallo} {et~al.}(2012){Gallo}, {Miller}, \& {Fender}}]{gallo12a}
{Gallo}, E., {Miller}, B.~P., \& {Fender}, R. 2012, \mnras, 423, 590

\bibitem[{{Gallo} {et~al.}(2014){Gallo}, {Miller-Jones}, {Russell}, {Jonker},
  {Homan}, {Plotkin}, {Markoff}, {Miller}, {Corbel}, \& {Fender}}]{gallo14a}
{Gallo}, E., {Miller-Jones}, J.~C.~A., {Russell}, D.~M., {et~al.} 2014, \mnras,
  445, 290

\bibitem[{{Galloway} \& {Cumming}(2006)}]{galloway06a}
{Galloway}, D.~K., \& {Cumming}, A. 2006, \apj, 652, 559

\bibitem[{{Greisen}(2003)}]{greisen03a}
{Greisen}, E.~W. 2003, Information Handling in Astronomy - Historical Vistas,
  285, 109

\bibitem[{{Halpern} {et~al.}(2013){Halpern}, {Gaidos}, {Sheffield},
  {Price-Whelan}, \& {Bogdanov}}]{halpern13a}
{Halpern}, J.~P., {Gaidos}, E., {Sheffield}, A., {Price-Whelan}, A.~M., \&
  {Bogdanov}, S. 2013, The Astronomer's Telegram, 5514, 1

\bibitem[{{Hynes} {et~al.}(2009){Hynes}, {Bradley}, {Rupen}, {Gallo}, {Fender},
  {Casares}, \& {Zurita}}]{hynes09a}
{Hynes}, R.~I., {Bradley}, C.~K., {Rupen}, M., {et~al.} 2009, \mnras, 399, 2239

\bibitem[{{Illarionov} \& {Sunyaev}(1975)}]{illarionov75a}
{Illarionov}, A.~F., \& {Sunyaev}, R.~A. 1975, \aap, 39, 185

\bibitem[{{Jones} \& {Ellison}(1991)}]{jones91a}
{Jones}, F.~C., \& {Ellison}, D.~C. 1991, \ssr, 58, 259

\bibitem[{{Jonker} {et~al.}(2012){Jonker}, {Miller-Jones}, {Homan}, {Tomsick},
  {Fender}, {Kaaret}, {Markoff}, \& {Gallo}}]{jonker12a}
{Jonker}, P.~G., {Miller-Jones}, J.~C.~A., {Homan}, J., {et~al.} 2012, \mnras,
  423, 3308

\bibitem[{{Kettenis} {et~al.}(2006){Kettenis}, {van Langevelde}, {Reynolds}, \&
  {Cotton}}]{kettenis06a}
{Kettenis}, M., {van Langevelde}, H.~J., {Reynolds}, C., \& {Cotton}, B. 2006,
  in Astronomical Society of the Pacific Conference Series, Vol. 351,
  Astronomical Data Analysis Software and Systems XV, ed. C.~{Gabriel},
  C.~{Arviset}, D.~{Ponz}, \& S.~{Enrique}, 497

\bibitem[{{Kreidberg} {et~al.}(2012){Kreidberg}, {Bailyn}, {Farr}, \&
  {Kalogera}}]{kreidberg12a}
{Kreidberg}, L., {Bailyn}, C.~D., {Farr}, W.~M., \& {Kalogera}, V. 2012, \apj,
  757, 36

\bibitem[{{Lane} {et~al.}(2014){Lane}, {Cotton}, {van Velzen}, {Clarke},
  {Kassim}, {Helmboldt}, {Lazio}, \& {Cohen}}]{lane14a}
{Lane}, W.~M., {Cotton}, W.~D., {van Velzen}, S., {et~al.} 2014, \mnras, 440,
  327

\bibitem[{{Linares} {et~al.}(2014){Linares}, {Bahramian}, {Heinke}, {Wijnands},
  {Patruno}, {Altamirano}, {Homan}, {Bogdanov}, \& {Pooley}}]{linares14a}
{Linares}, M., {Bahramian}, A., {Heinke}, C., {et~al.} 2014, \mnras, 438, 251

\bibitem[{{MacDonald} {et~al.}(2014){MacDonald}, {Bailyn}, {Buxton},
  {Cantrell}, {Chatterjee}, {Kennedy-Shaffer}, {Orosz}, {Markwardt}, \&
  {Swank}}]{macdonald14a}
{MacDonald}, R.~K.~D., {Bailyn}, C.~D., {Buxton}, M., {et~al.} 2014, \apj, 784,
  2

\bibitem[{{Markoff} {et~al.}(2001){Markoff}, {Falcke}, \&
  {Fender}}]{markoff01a}
{Markoff}, S., {Falcke}, H., \& {Fender}, R. 2001, \aap, 372, L25

\bibitem[{{McMullin} {et~al.}(2007){McMullin}, {Waters}, {Schiebel}, {Young},
  \& {Golap}}]{mcmullin07a}
{McMullin}, J.~P., {Waters}, B., {Schiebel}, D., {Young}, W., \& {Golap}, K.
  2007, in Astronomical Society of the Pacific Conference Series, Vol. 376,
  Astronomical Data Analysis Software and Systems XVI, ed. R.~A. {Shaw},
  F.~{Hill}, \& D.~J. {Bell}, 127

\bibitem[{{Merloni} {et~al.}(2003){Merloni}, {Heinz}, \& {di
  Matteo}}]{merloni03a}
{Merloni}, A., {Heinz}, S., \& {di Matteo}, T. 2003, \mnras, 345, 1057

\bibitem[{{Migliari} \& {Fender}(2006)}]{migliari06a}
{Migliari}, S., \& {Fender}, R.~P. 2006, \mnras, 366, 79

\bibitem[{{Migliari} {et~al.}(2003){Migliari}, {Fender}, {Rupen}, {Jonker},
  {Klein-Wolt}, {Hjellming}, \& {van der Klis}}]{migliari03a}
{Migliari}, S., {Fender}, R.~P., {Rupen}, M., {et~al.} 2003, \mnras, 342, L67

\bibitem[{{Migliari} {et~al.}(2011){Migliari}, {Miller-Jones}, \&
  {Russell}}]{migliari11a}
{Migliari}, S., {Miller-Jones}, J.~C.~A., \& {Russell}, D.~M. 2011, \mnras,
  415, 2407

\bibitem[{{Miller-Jones} {et~al.}(2008){Miller-Jones}, {Gallo}, {Rupen},
  {Mioduszewski}, {Brisken}, {Fender}, {Jonker}, \&
  {Maccarone}}]{miller-jones08a}
{Miller-Jones}, J.~C.~A., {Gallo}, E., {Rupen}, M.~P., {et~al.} 2008, \mnras,
  388, 1751

\bibitem[{{Miller-Jones} {et~al.}(2010){Miller-Jones}, {Sivakoff},
  {Altamirano}, {Tudose}, {Migliari}, {Dhawan}, {Fender}, {Garrett}, {Heinz},
  {K{\"o}rding}, {Krimm}, {Linares}, {Maitra}, {Markoff}, {Paragi},
  {Remillard}, {Rupen}, {Rushton}, {Russell}, {Sarazin}, \&
  {Spencer}}]{miller-jones10a}
{Miller-Jones}, J.~C.~A., {Sivakoff}, G.~R., {Altamirano}, D., {et~al.} 2010,
  \apjl, 716, L109

\bibitem[{{Papitto} {et~al.}(2015){Papitto}, {de Martino}, {Belloni}, {Burgay},
  {Pellizzoni}, {Possenti}, \& {Torres}}]{papitto15a}
{Papitto}, A., {de Martino}, D., {Belloni}, T.~M., {et~al.} 2015, \mnras, 449,
  L26

\bibitem[{{Papitto} {et~al.}(2014){Papitto}, {Torres}, \& {Li}}]{papitto14a}
{Papitto}, A., {Torres}, D.~F., \& {Li}, J. 2014, \mnras, 438, 2105

\bibitem[{{Papitto} {et~al.}(2013){Papitto}, {Ferrigno}, {Bozzo}, {Rea},
  {Pavan}, {Burderi}, {Burgay}, {Campana}, {di Salvo}, {Falanga},
  {Filipovi{\'c}}, {Freire}, {Hessels}, {Possenti}, {Ransom}, {Riggio},
  {Romano}, {Sarkissian}, {Stairs}, {Stella}, {Torres}, {Wieringa}, \&
  {Wong}}]{papitto13a}
{Papitto}, A., {Ferrigno}, C., {Bozzo}, E., {et~al.} 2013, \nat, 501, 517

\bibitem[{{Patruno} \& {Watts}(2012)}]{patruno12a}
{Patruno}, A., \& {Watts}, A.~L. 2012, ArXiv e-prints, arXiv:1206.2727

\bibitem[{{Patruno} {et~al.}(2014){Patruno}, {Archibald}, {Hessels},
  {Bogdanov}, {Stappers}, {Bassa}, {Janssen}, {Kaspi}, {Tendulkar}, \&
  {Lyne}}]{patruno14a}
{Patruno}, A., {Archibald}, A.~M., {Hessels}, J.~W.~T., {et~al.} 2014, \apjl,
  781, L3

\bibitem[{{Pidopryhora} {et~al.}(2009){Pidopryhora}, {Keimpema}, \&
  {Kettenis}}]{pidopryhora09a}
{Pidopryhora}, Y., {Keimpema}, A., \& {Kettenis}, M. 2009, in 8th International
  e-VLBI Workshop

\bibitem[{{Plotkin} {et~al.}(2012){Plotkin}, {Markoff}, {Kelly}, {K{\"o}rding},
  \& {Anderson}}]{plotkin12a}
{Plotkin}, R.~M., {Markoff}, S., {Kelly}, B.~C., {K{\"o}rding}, E., \&
  {Anderson}, S.~F. 2012, \mnras, 419, 267

\bibitem[{{Radhakrishnan} \& {Srinivasan}(1982)}]{radhakrishnan82a}
{Radhakrishnan}, V., \& {Srinivasan}, G. 1982, Current Science, 51, 1096

\bibitem[{{Ratti} {et~al.}(2012){Ratti}, {Jonker}, {Miller-Jones}, {Torres},
  {Homan}, {Markoff}, {Tomsick}, {Kaaret}, {Wijnands}, {Gallo}, {{\"O}zel},
  {Steeghs}, \& {Fender}}]{ratti12a}
{Ratti}, E.~M., {Jonker}, P.~G., {Miller-Jones}, J.~C.~A., {et~al.} 2012,
  \mnras, 423, 2656

\bibitem[{{Reynolds} {et~al.}(2002){Reynolds}, {Paragi}, \&
  {Garrett}}]{reynolds02a}
{Reynolds}, C., {Paragi}, Z., \& {Garrett}, M. 2002, ArXiv Astrophysics
  e-prints, astro-ph/0205118

\bibitem[{{Romanova} {et~al.}(2005){Romanova}, {Ustyugova}, {Koldoba}, \&
  {Lovelace}}]{romanova05a}
{Romanova}, M.~M., {Ustyugova}, G.~V., {Koldoba}, A.~V., \& {Lovelace},
  R.~V.~E. 2005, \apjl, 635, L165

\bibitem[{{Romanova} {et~al.}(2009){Romanova}, {Ustyugova}, {Koldoba}, \&
  {Lovelace}}]{romanova09a}
---. 2009, \mnras, 399, 1802

\bibitem[{{Roy} {et~al.}(2015){Roy}, {Ray}, {Bhattacharyya}, {Stappers},
  {Chengalur}, {Deneva}, {Camilo}, {Johnson}, {Wolff}, {Hessels}, {Bassa},
  {Keane}, {Ferrara}, {Harding}, \& {Wood}}]{roy15a}
{Roy}, J., {Ray}, P.~S., {Bhattacharyya}, B., {et~al.} 2015, \apjl, 800, L12

\bibitem[{{Rupen} {et~al.}(2005){Rupen}, {Dhawan}, \&
  {Mioduszewski}}]{rupen05a}
{Rupen}, M.~P., {Dhawan}, V., \& {Mioduszewski}, A.~J. 2005, The Astronomer's
  Telegram, 524, 1

\bibitem[{{Russell} {et~al.}(2007){Russell}, {Fender}, \&
  {Jonker}}]{russell07a}
{Russell}, D.~M., {Fender}, R.~P., \& {Jonker}, P.~G. 2007, \mnras, 379, 1108

\bibitem[{{Russell} \& {Shahbaz}(2014)}]{russell14a}
{Russell}, D.~M., \& {Shahbaz}, T. 2014, \mnras, 438, 2083

\bibitem[{{Russell} {et~al.}(2012){Russell}, {Curran}, {Mu{\~n}oz-Darias},
  {Lewis}, {Motta}, {Stiele}, {Belloni}, {Miller-Jones}, {Jonker}, {O'Brien},
  {Homan}, {Casella}, {Gandhi}, {Soleri}, {Markoff}, {Maitra}, {Gallo}, \&
  {Cadolle Bel}}]{russell12a}
{Russell}, D.~M., {Curran}, P.~A., {Mu{\~n}oz-Darias}, T., {et~al.} 2012,
  \mnras, 419, 1740

\bibitem[{{Russell} {et~al.}(2013){Russell}, {Markoff}, {Casella}, {Cantrell},
  {Chatterjee}, {Fender}, {Gallo}, {Gandhi}, {Homan}, {Maitra}, {Miller-Jones},
  {O'Brien}, \& {Shahbaz}}]{russell13a}
{Russell}, D.~M., {Markoff}, S., {Casella}, P., {et~al.} 2013, \mnras, 429, 815

\bibitem[{{Rutledge} {et~al.}(2002){Rutledge}, {Bildsten}, {Brown}, {Pavlov},
  \& {Zavlin}}]{rutledge02a}
{Rutledge}, R.~E., {Bildsten}, L., {Brown}, E.~F., {Pavlov}, G.~G., \&
  {Zavlin}, V.~E. 2002, \apj, 577, 346

\bibitem[{{Shepherd} {et~al.}(1994){Shepherd}, {Pearson}, \&
  {Taylor}}]{shepherd94a}
{Shepherd}, M.~C., {Pearson}, T.~J., \& {Taylor}, G.~B. 1994, in Bulletin of
  the American Astronomical Society, Vol.~26, Bulletin of the American
  Astronomical Society, 987--989

\bibitem[{{Smarr} \& {Blandford}(1976)}]{smarr76a}
{Smarr}, L.~L., \& {Blandford}, R. 1976, \apj, 207, 574

\bibitem[{{Soleri} {et~al.}(2010){Soleri}, {Fender}, {Tudose}, {Maitra},
  {Bell}, {Linares}, {Altamirano}, {Wijnands}, {Belloni}, {Casella},
  {Miller-Jones}, {Muxlow}, {Klein-Wolt}, {Garrett}, \& {van der
  Klis}}]{soleri10a}
{Soleri}, P., {Fender}, R., {Tudose}, V., {et~al.} 2010, \mnras, 406, 1471

\bibitem[{{Spencer} {et~al.}(2013){Spencer}, {Rushton},
  {Ba{\l}uci{\'n}ska-Church}, {Paragi}, {Schulz}, {Wilms}, {Pooley}, \&
  {Church}}]{spencer13a}
{Spencer}, R.~E., {Rushton}, A.~P., {Ba{\l}uci{\'n}ska-Church}, M., {et~al.}
  2013, \mnras, 435, L48

\bibitem[{{Spruit} \& {Taam}(1993)}]{spruit93a}
{Spruit}, H.~C., \& {Taam}, R.~E. 1993, \apj, 402, 593

\bibitem[{{Stappers} {et~al.}(2011){Stappers}, {Hessels}, {Alexov}, {Anderson},
  {Coenen}, {Hassall}, {Karastergiou}, {Kondratiev}, {Kramer}, {van Leeuwen},
  {Mol}, {Noutsos}, {Romein}, {Weltevrede}, {Fender}, {Wijers}, {B{\"a}hren},
  {Bell}, {Broderick}, {Daw}, {Dhillon}, {Eisl{\"o}ffel}, {Falcke},
  {Griessmeier}, {Law}, {Markoff}, {Miller-Jones}, {Scheers}, {Spreeuw},
  {Swinbank}, {Ter Veen}, {Wise}, {Wucknitz}, {Zarka}, {Anderson}, {Asgekar},
  {Avruch}, {Beck}, {Bennema}, {Bentum}, {Best}, {Bregman}, {Brentjens}, {van
  de Brink}, {Broekema}, {Brouw}, {Br{\"u}ggen}, {de Bruyn}, {Butcher},
  {Ciardi}, {Conway}, {Dettmar}, {van Duin}, {van Enst}, {Garrett}, {Gerbers},
  {Grit}, {Gunst}, {van Haarlem}, {Hamaker}, {Heald}, {Hoeft}, {Holties},
  {Horneffer}, {Koopmans}, {Kuper}, {Loose}, {Maat}, {McKay-Bukowski},
  {McKean}, {Miley}, {Morganti}, {Nijboer}, {Noordam}, {Norden}, {Olofsson},
  {Pandey-Pommier}, {Polatidis}, {Reich}, {R{\"o}ttgering}, {Schoenmakers},
  {Sluman}, {Smirnov}, {Steinmetz}, {Sterks}, {Tagger}, {Tang}, {Vermeulen},
  {Vermaas}, {Vogt}, {de Vos}, {Wijnholds}, {Yatawatta}, \&
  {Zensus}}]{stappers11a}
{Stappers}, B.~W., {Hessels}, J.~W.~T., {Alexov}, A., {et~al.} 2011, \aap, 530,
  A80

\bibitem[{{Stappers} {et~al.}(2014){Stappers}, {Archibald}, {Hessels}, {Bassa},
  {Bogdanov}, {Janssen}, {Kaspi}, {Lyne}, {Patruno}, {Tendulkar}, {Hill}, \&
  {Glanzman}}]{stappers14a}
{Stappers}, B.~W., {Archibald}, A.~M., {Hessels}, J.~W.~T., {et~al.} 2014,
  \apj, 790, 39

\bibitem[{{Stirling} {et~al.}(2001){Stirling}, {Spencer}, {de la Force},
  {Garrett}, {Fender}, \& {Ogley}}]{stirling01a}
{Stirling}, A.~M., {Spencer}, R.~E., {de la Force}, C.~J., {et~al.} 2001,
  \mnras, 327, 1273

\bibitem[{{Strader} {et~al.}(2012){Strader}, {Chomiuk}, {Maccarone},
  {Miller-Jones}, \& {Seth}}]{strader12a}
{Strader}, J., {Chomiuk}, L., {Maccarone}, T.~J., {Miller-Jones}, J.~C.~A., \&
  {Seth}, A.~C. 2012, \nat, 490, 71

\bibitem[{{Syunyaev} \& {Shakura}(1977)}]{syunyaev77a}
{Syunyaev}, R.~A., \& {Shakura}, N.~I. 1977, Soviet Astronomy Letters, 3, 138

\bibitem[{{Tendulkar} {et~al.}(2014){Tendulkar}, {Yang}, {An}, {Kaspi},
  {Archibald}, {Bassa}, {Bellm}, {Bogdanov}, {Harrison}, {Hessels}, {Janssen},
  {Lyne}, {Patruno}, {Stappers}, {Stern}, {Tomsick}, {Boggs}, {Chakrabarty},
  {Christensen}, {Craig}, {Hailey}, \& {Zhang}}]{tendulkar14a}
{Tendulkar}, S.~P., {Yang}, C., {An}, H., {et~al.} 2014, \apj, 791, 77

\bibitem[{{Thorstensen} \& {Armstrong}(2005)}]{thorstensen05a}
{Thorstensen}, J.~R., \& {Armstrong}, E. 2005, \aj, 130, 759

\bibitem[{{Tudose} {et~al.}(2009){Tudose}, {Fender}, {Linares}, {Maitra}, \&
  {van der Klis}}]{tudose09a}
{Tudose}, V., {Fender}, R.~P., {Linares}, M., {Maitra}, D., \& {van der Klis},
  M. 2009, \mnras, 400, 2111

\bibitem[{{Ustyugova} {et~al.}(2006){Ustyugova}, {Koldoba}, {Romanova}, \&
  {Lovelace}}]{ustyugova06a}
{Ustyugova}, G.~V., {Koldoba}, A.~V., {Romanova}, M.~M., \& {Lovelace},
  R.~V.~E. 2006, \apj, 646, 304

\bibitem[{{van Haarlem} {et~al.}(2013){van Haarlem}, {Wise}, {Gunst}, {Heald},
  {McKean}, {Hessels}, {de Bruyn}, {Nijboer}, {Swinbank}, {Fallows},
  {Brentjens}, {Nelles}, {Beck}, {Falcke}, {Fender}, {H{\"o}randel},
  {Koopmans}, {Mann}, {Miley}, {R{\"o}ttgering}, {Stappers}, {Wijers},
  {Zaroubi}, {van den Akker}, {Alexov}, {Anderson}, {Anderson}, {van Ardenne},
  {Arts}, {Asgekar}, {Avruch}, {Batejat}, {B{\"a}hren}, {Bell}, {Bell}, {van
  Bemmel}, {Bennema}, {Bentum}, {Bernardi}, {Best}, {B{\^\i}rzan}, {Bonafede},
  {Boonstra}, {Braun}, {Bregman}, {Breitling}, {van de Brink}, {Broderick},
  {Broekema}, {Brouw}, {Br{\"u}ggen}, {Butcher}, {van Cappellen}, {Ciardi},
  {Coenen}, {Conway}, {Coolen}, {Corstanje}, {Damstra}, {Davies}, {Deller},
  {Dettmar}, {van Diepen}, {Dijkstra}, {Donker}, {Doorduin}, {Dromer}, {Drost},
  {van Duin}, {Eisl{\"o}ffel}, {van Enst}, {Ferrari}, {Frieswijk}, {Gankema},
  {Garrett}, {de Gasperin}, {Gerbers}, {de Geus}, {Grie{\ss}meier}, {Grit},
  {Gruppen}, {Hamaker}, {Hassall}, {Hoeft}, {Holties}, {Horneffer}, {van der
  Horst}, {van Houwelingen}, {Huijgen}, {Iacobelli}, {Intema}, {Jackson},
  {Jelic}, {de Jong}, {Juette}, {Kant}, {Karastergiou}, {Koers}, {Kollen},
  {Kondratiev}, {Kooistra}, {Koopman}, {Koster}, {Kuniyoshi}, {Kramer},
  {Kuper}, {Lambropoulos}, {Law}, {van Leeuwen}, {Lemaitre}, {Loose}, {Maat},
  {Macario}, {Markoff}, {Masters}, {McFadden}, {McKay-Bukowski}, {Meijering},
  {Meulman}, {Mevius}, {Middelberg}, {Millenaar}, {Miller-Jones}, {Mohan},
  {Mol}, {Morawietz}, {Morganti}, {Mulcahy}, {Mulder}, {Munk}, {Nieuwenhuis},
  {van Nieuwpoort}, {Noordam}, {Norden}, {Noutsos}, {Offringa}, {Olofsson},
  {Omar}, {Orr{\'u}}, {Overeem}, {Paas}, {Pandey-Pommier}, {Pandey}, {Pizzo},
  {Polatidis}, {Rafferty}, {Rawlings}, {Reich}, {de Reijer}, {Reitsma},
  {Renting}, {Riemers}, {Rol}, {Romein}, {Roosjen}, {Ruiter}, {Scaife}, {van
  der Schaaf}, {Scheers}, {Schellart}, {Schoenmakers}, {Schoonderbeek},
  {Serylak}, {Shulevski}, {Sluman}, {Smirnov}, {Sobey}, {Spreeuw}, {Steinmetz},
  {Sterks}, {Stiepel}, {Stuurwold}, {Tagger}, {Tang}, {Tasse}, {Thomas},
  {Thoudam}, {Toribio}, {van der Tol}, {Usov}, {van Veelen}, {van der Veen},
  {ter Veen}, {Verbiest}, {Vermeulen}, {Vermaas}, {Vocks}, {Vogt}, {de Vos},
  {van der Wal}, {van Weeren}, {Weggemans}, {Weltevrede}, {White}, {Wijnholds},
  {Wilhelmsson}, {Wucknitz}, {Yatawatta}, {Zarka}, {Zensus}, \& {van
  Zwieten}}]{van-haarlem13a}
{van Haarlem}, M.~P., {Wise}, M.~W., {Gunst}, A.~W., {et~al.} 2013, \aap, 556,
  A2

\bibitem[{{Wijnands} {et~al.}(2005){Wijnands}, {Heinke}, {Pooley}, {Edmonds},
  {Lewin}, {Grindlay}, {Jonker}, \& {Miller}}]{wijnands05a}
{Wijnands}, R., {Heinke}, C.~O., {Pooley}, D., {et~al.} 2005, \apj, 618, 883

\bibitem[{{Wijnands} \& {van der Klis}(1998)}]{wijnands98a}
{Wijnands}, R., \& {van der Klis}, M. 1998, \nat, 394, 344

\end{thebibliography}

\end{document}